# Spatial Colorimetric Mapping of pH and Dissolved Carbon during Convective Dissolution of $CO_2$: Constraints from Domain, Geometry, and Indicator Mix on Measurement Reliability

Joe-sam Nkuah[1], Vance Sisson[2], Haochen Li[1], Anna Herring[1]
1. Department of Civil and Environmental Engineering, University of Tennessee–Knoxville; TN, USA
2. Department of Chemical Engineering, University of Tennessee–Knoxville; TN, USA.

## Abstract

Quantifying the dissolution and buoyancy-driven transport of $CO_2$-rich water is critical to understanding solubility trapping mechanism in geologic storage. Analog-fluid density driven convection experiments do not represent $CO_2$ dissolution chemistry or the aqueous $CO_2$ density-concentration monotonic relationship. X-ray tomography does not provide enough contrast to directly measure dissolved $CO_2$. Spatial colorimetry provides an inexpensive method to map pH and estimate total dissolved inorganic carbon ($C_T$) via carbonate equilibrium assumptions. However, the effects of experimental conditions such as domain (free fluid and porous), geometry (2D and 3D), and indicator mix concentration on the reliability of pH and $C_T$ measurements remain unexplored. We evaluated how these factors affect the reliability of measurements using two different colorimetric models (hue and spherical coordinates (θ and ϕ)). Hue was reliable in all domains whereas θ–ϕ was reliable in free fluid. Lower indicator mix concentration increased the chromatic sensitivity at low pH in free fluid; however, it did not produce sufficient color contrast for pH calibration and mapping in the porous domain. In the porous domain, θ–ϕ calibration was 33 times more susceptible to noise than hue. Temporal pH noise was less than 0.01 units; spatial noise changed domain-mean $C_T$ by 0.6-4.4%. Both models were able to resolve $CO_2$-rich plumes in free fluid; however, θ–ϕ had sharper plume boundaries than hue. The temporal evolution of $CO_2$ dissolution resolved by the color models agreed more in 2D than 3D. These findings provide a reliability framework for studying subsurface dissolution and mixing of $CO_2$.

## 1 Introduction

Geologic carbon storage (GCS) has proven to be a feasible method to reduce global anthropogenic carbon dioxide ($CO_2$) emissions; about 40 gigatons could be removed globally by mid-century at effective costs if GCS is coupled with other carbon-negative emissions technologies (Fawzy et al., 2020; Hameli et al., 2022). Effective containment of $CO_2$ requires understanding of subsurface flow chemistry and physics, and the impact of processes such as convective dissolution and mixing. When $CO_2$ is injected into a geologic reservoir (which could include saline aquifers or depleted oil and gas reservoirs), the reservoir-level temperature and pressure trigger the formation of a supercritical $CO_2$ phase which is buoyant relative to the formation brine. Supercritical $CO_2$ buoyantly migrates upwards, accumulating beneath low-permeability caprocks. Over time, $CO_2$ dissolves into the underlying brine, increasing its density and leading to gravitational instabilities (Thomas et al., 2015; De N et al., 2021; Emami-Meybodi et al., 2015). The resulting $CO_2$-rich, dense fluid sinks in fingers or plume morphologies as "fresh," non-$CO_2$-laden water is pulled upward, thereby enhancing $CO_2$ dissolution and mixing far beyond molecular diffusion. This process is known as convective dissolution of $CO_2$, and it enhances solubility trapping, one of the four trapping mechanisms (structural, residual, or capillary, solubility, and mineral trapping) that ensure that $CO_2$ remains stored on geologic time frame (IPCC, 2005).

Convective dissolution accelerates the downward transport and mixing of dissolved inorganic carbon, improving storage efficiency and impacting key flow and mixing metrics such as dissolution flux, mixing length, and onset time of instability; all of which control the rate and extent of $CO_2$–brine interaction and subsequent geochemical reactions (Ko et al., 2020; De et al., 2021; Emami-Meybodi et al., 2015).

Recent studies on $CO_2$ convective dissolution include direct numerical simulations (Benhammadi R et al., 2025; Hidalgo et al., 2012), Hele-Shaw experiments (J.A. Letelier et al 2023; De Paoli, M., et al. 2023), packed-bead experiments (Brouzet et al., 2022; Mykkeltvedt T.S. et al., 2025), and 3D porous-media imaging (Liyanage, R. et al., 2024). These studies often use analog fluids in combination with dimensionless analysis to measure how porous media parameters and boundary conditions affect dissolution onset time, finger growth and propagation, plume interaction, and dissolution flux. Analog $CO_2$ fluids (ethylene and methanol glycol) provide controlled density contrasts and clear optical visualization but do not replicate the acid-base chemistry of $CO_2$ dissolution, nor the monotonic density-concentration relationship of aqueous $CO_2$ (Jafari Raad et al., 2016). Some analog fluid mixtures have large viscosity changes with composition which can alter the buoyancy-driven flow resistance and modify transport coefficients in a way that can be unlike $CO_2$–water/brine (Jafari Raad et al., 2016). Alternate visualization methods, e.g., X-ray Computed Tomography (CT) imaging, can provide good 3D visualization, but cannot directly resolve $CO_2$ concentration gradients, because the density increase associated with dissolved $CO_2$ (approximately 1-2%) does not produce sufficient X-ray contrast. Although image processing pipelines have been developed to estimate dissolved solute concentrations (R. Huang et al., 2023), these rely on measurements of gas-phase morphology dynamics; thus, X-ray CT experiments targeting convective dissolution have again generally relied on analog solutes.

To replicate pH evolution and $CO_2$/aqueous phase density relationships, as well as measurement of total dissolved inorganic carbon concentration ($C_T$) from carbonate equilibrium chemistry, non-analog fluid experimental methodologies needs to be investigated. Spatial colorimetric pH mapping offers a comparatively simple and low-cost approach for visualizing $CO_2$ dissolution resulting from carbonate equilibrium chemistry in an optically accessible flow cells: pH-dependent color changes of an indicator solution are converted pixel-wise into pH maps, enabling visualization and quantification of the spatial and temporal evolution of $CO_2$-rich convective plumes. Such methods have been deployed to visualize $CO_2$ plume propagation and finger morphology in Hele-Shaw cells (Birgisson et al. 2025, Xu et al. 2026). A key advantage of this method is that the pixel-wise pH fields resolved from pure $CO_2$ dissolution chemistry can be converted into total dissolved inorganic carbon fields using carbonate equilibrium speciation assumptions or water chemistry models. This is particularly informative for studies on convective dissolution because colorimetric imaging provides real-time visualization of $CO_2$-rich fingers moving downward through the fluid-filled medium, allowing direct measurement of the fluid structures that control mixing and dissolution mass flux. This is in contrast to bulk gas-pressure experiments, which provide only integrated concentration measurement and obscure the flow physics.

Recently, Birgisson et al. (2025) demonstrated the utility of colorimetric mapping by measuring both pH and total dissolved inorganic carbon ($C_T$) during density-driven $CO_2$ convection in water in a quasi-2D Hele-Shaw cell. A three-indicator mixture consisting of methyl red (MR), bromothymol blue (BTB), and thymol blue (TB), yielded a broad pH-sensitive color response from pH 4.0 to 9.5; RGB vectors were transformed via angular color representation into 3D space spherical coordinates (polar ($\theta$) and azimuthal ($\phi$) angles) to reduce sensitivity to brightness variations. The calibrated angular color path was then used to assign pH values to each image pixel, and the pH field was converted to total dissolved carbon using carbonate equilibrium relationships. They also evaluated how pH uncertainty propagates into $C_{T,}$ showing that addition of normally distributed pH noise with a standard deviation of 0.10 pH units changed the estimated total dissolved carbon by only about 2.48% on average; a larger pH uncertainty of 0.20 pH units produced an approximately 10% change in total carbon.

Xu et al. (2026) expanded this approach by using the same mix indicator as Birggison et al. (2025) and systematically comparing three different colorimetric quantification approaches: hue in HSV color space, gray-difference (assuming the Beer-Lambert law), and angular RGB representation using $\phi$ –$\theta$ coordinates. The flow cell used was milled transparent polymethyl methacrylate (PMMA) with uniformly distributed cylindrical posts representing a porous medium. They concluded that the hue-based quantification was the most robust, showing stronger invariance to fluid thickness, camera settings, and LED luminance. In contrast, the gray-difference and $\theta$-$\phi$ methods were more sensitive to optical and environmental variations, and the pH interpolation error was high when the calibration path became non-monotonic. However, despite the hue-based method's relatively high performance, the low pH region of the calibration curve ($4 < pH < 6$) exhibited reduced color sensitivity.

To date, the effects of some important experimental variables on spatial colorimetry-based pH and carbon quantification remain unexplored: the effect of domain (i.e., free fluid vs. porous media), flow cell geometry (2D vs. 3D), and indicator mix concentration. Herein, we analyze how these factors affect the reliability of measurements of resolved pH and $C_T$ using two colorimetric models: the hue model and the ϕ–θ model of Birgisson et al. (2025).

The hue method is based on the hue-saturation -value (HSV) representation of color space, where one coordinate, hue (i.e., angle on the color wheel) represents color. In contrast, the ϕ–θ model transformation stores two coordinates (polar (θ) and azimuthal (ϕ) angles) in 3D angular space (Birgisson et al., 2025). Herein, we quantify the reliability of spatial colorimetric mapping of pH and $C_T$ via these two color representations during convective dissolution of $CO_2$, considering the different flow domains (i.e., whether porous media is present or not), flow cell geometries, and different indicator mix concentrations. We provide a constraint-reliability framework for hue- and θ-ϕ-based pH and $C_T$ mapping across free-fluid 2D (FF 2D), free-fluid 3D (FF 3D), and porous domains.

## 2 Experimental methods

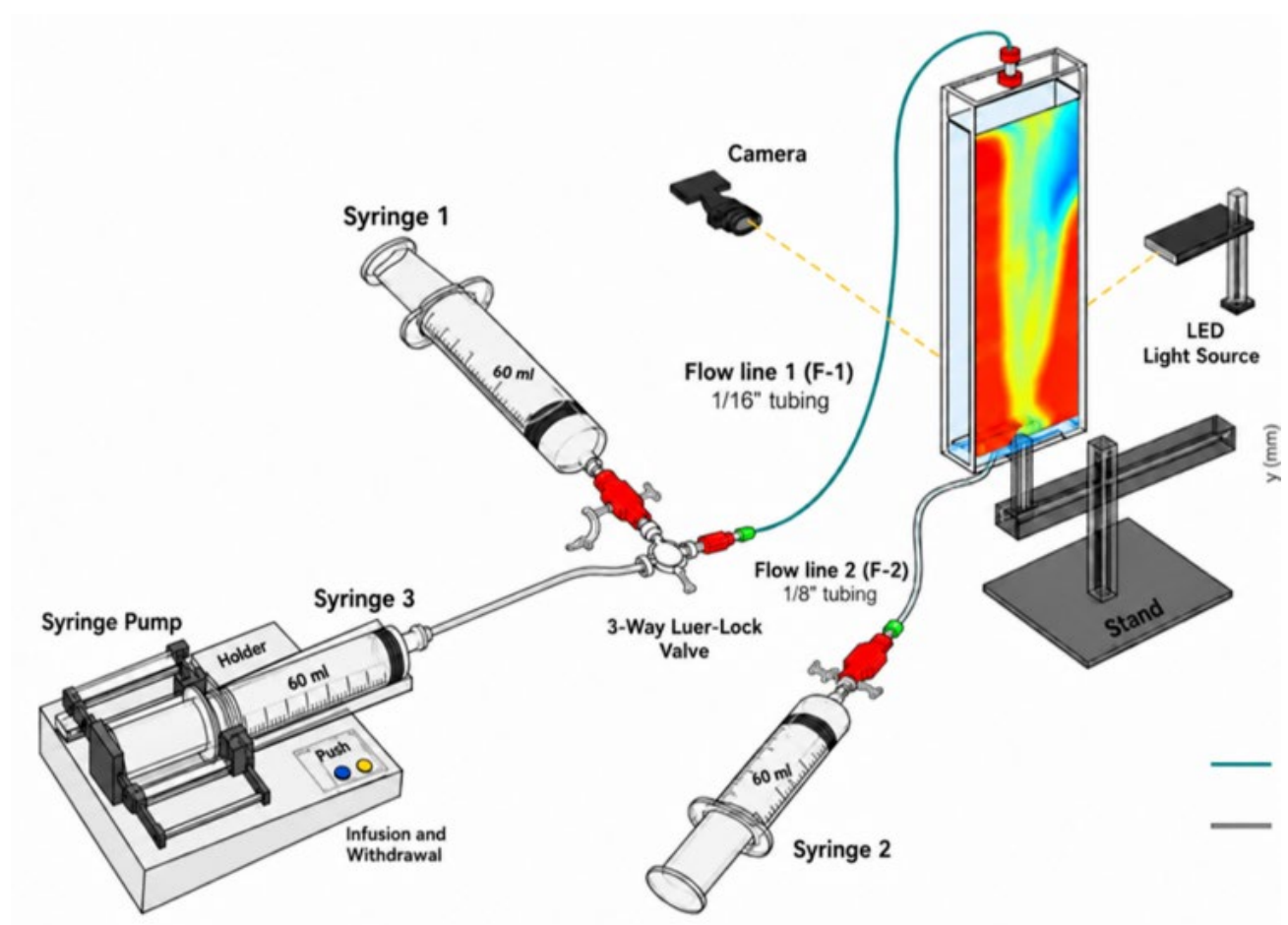

Fig. 1: Experimental setup

### 2.1 Experimental Setup

The setup included a Blackfly camera (Teledyne Flir Blackfly, S-8.9MP CMOS Sensor), LED light source, Harvard PHD syringe pump, and three custom-built transparent column flow cells (2D, quasi-2D, and 3D geometries) made with polycarbonate sheets (see Table 1). Cell tops were sealed with a Teflon gasket (2D) or a rubber stopper (3D); total cell height was 100 mm for all configurations, with a 10 mm headspace maintained above the fluid interface for $CO_2$ injection. Flow cells were connected to a 60 mL gas-filled syringe 3 installed in a Harvard PHD syringe pump via a 1/16" PEEK tubing flow line from the top, and to a 60 mL syringe 2 by 1/8" PEEK tubing for refilling the cell with indicator solution from the bottom (Fig.

1). 2D and 3D cells were filled with indicator solution to achieve a “free fluid” domain. To create a porous medium system, the quasi-2D cell was packed with 1 mm-diameter, 1.20 mm-thick Teflon ™ FEP pellets with a refractive index (RI) similar to water (RI = 1.34) prior to saturation with the indicator solution. $CO_2$ gas (>99.5% purity) was injected continuously into the headspace at a flow rate of 0.75 ml/min for 50 minutes at ambient temperature and pressure. The interaction of $CO_2$ with the indicator solution during and after the injection was captured using an acquisition framerate of 1 Hz.

Table 1: Experimental conditions

| Exp | Flow Cell | Dimension (mm) | Domain | Conc. Mix | Permeability (k, $m^2$) and porosity ($\phi_p$) |
|---|---|---|---|---|---|
| 1 | 2D | 100 x 18 x 0.95 | Free fluid | 1& 2 | k= $7.5 \times 10^{-8}$<br>$\phi_p$ = 1.0 |
| 2 | Quasi-2D | 100 x 18 x 1.75 | Porous Medium | 2 | k= $2.6\times10^{-10}$<br>$\phi_p$ = 0.29 |
| 3 | 3D | 100 x 22 x 10 | Free fluid | 1& 2 | k= $8.3 \times 10^{-6}$<br>$\phi_p$ = 1.0 |

**2.2 Indicator Mix Preparation**

To achieve a continuous color spectrum for measuring pH and $C_T$, three colorimetric indicators purchased from RICCA Chemicals (see Table 2) were mixed, and the pH was measured using a standard pH probe. Alkalinity was estimated via the PHREEQC equilibrium solver (sec 2.7.1). pH buffers were used to establish a wide range of pH calibration solutions ranging from pH 3.5 to 8. The three indicators were combined to enable discrimination over the full pH range (Birggison et al., 2025).

Table 2: Colorimetric indicator concentrations and water chemistry parameters for the two mixes investigated.

| Mix | pH Indicators Concentration (mg/l) | | | Alkalinity (mol/l) | Initial pH |
|---|---|---|---|---|---|
| | BTB | TB | MR | | |
| 1 | 13.2 | 13.2 | 6.61 | $2.8 \times 10^{-5}$ | 6.4 - 6.6 |
| 2 | 50 | 25 | 32 | $1.2 \times 10^{-4}$ | 7.2 -7.4 |

Two different mix concentrations were investigated (see Table 2) to assess the effect of dye loading on chromatic compression in the low-pH range (5.5 - 4.0) of the pH calibration curve, a problem previously identified by Xu et al. 2026 (section S3.0). The comparison of calibration curves derived from the two mixes showed that the lower concentration mix 1 produced better separation at low pH values (vs. mix 2) in both free-fluid geometries (Tables S1-S3 and Figs. S3-S7). However, mix 1 did not produce adequate color contrast for the porous medium calibration and mapping. To compare the porous and free-fluid domains with the same solution chemistry, mix 2 was used for all domain $CO_2$ dissolution experiments and the subsequent measurement reliability test.

### 2.3 Image Preprocessing

Image data was pre-processed in MATLAB to address non-uniform lighting, background intensity gradients, and artifacts associated with LED lighting. The first frame of the time-resolved stack (i.e., before $CO_2$ injection) was used as a brightfield frame to compute two correction maps: flat field and gradient correction. The flat-field correction is performed using a Gaussian blur filter on the grayscale version of the image. This compensates for illumination non-uniformity (the image is generally brighter at the center and darker at the edges). The resulting blurred image is used to normalize the post-$CO_2$ injection RGB frames on a per-pixel basis. A 2D polynomial gradient correction is then applied by fitting a second-order polynomial surface to the grayscale image of the flat-field-corrected frame. A least-squares solution is used to estimate the polynomial coefficients, and the fitted surface is then used to further normalize images. The two corrections (flat-field and polynomial) were applied to all subsequent experimental images captured (see Fig. S1).

### 2.4 Pore map segmentation

For the porous domain, pore segmentation was performed on a raw RGB image of the packed Teflon cell. The image was converted to grayscale and normalized to a 0–1 intensity scale. Direct threshold segmentation was applied; pixels with intensity $\leq 0.60$ were classified as pore space, while pixels with intensity $\geq 0.64$ were classified as solid Teflon grains. Pixels with intensities between 0.60 and 0.64 were treated as unresolved transition pixels, mainly associated with Teflon edges. For pH and $C_T$ mapping, a more conservative pore mask was applied by removing pore pixels within 2 pixels of the nearest labeled solid boundary. This final pH mapping mask corresponded to 29.65% of the image area (see Fig. S2).

### 2.5 pH calibration and mapping

During initial calibration, the flow cells were iteratively filled with the calibration solutions ranging from pH 3.5–8, and their respective images were acquired. Digital images store color as three values: red (R), green (G), and blue (B); each of the raw calibration images taken were split into separate RGB channels, and the mean intensity of the three channels was extracted and fed into the two calibration models (hue and $\theta$–$\phi$). The spherical coordinate ($\theta$–$\phi$) pH model transforms extracted RGB intensity values to a 3D vector in space, normalizes it, and converts it into angular coordinates: polar angle ($\theta$), azimuthal angle ($\phi$), and radial distance (P) (eq. 1-3). The radial distance represents the magnitude length of the color vector, while $\theta$ accounts for the blue color shift and $\phi$ accounts for the color transition between red and green channels in space. From calibration images, a 2D scattered interpolant pH = f(θ, ϕ) was constructed and evaluated pixel-wise to relate θ and ϕ to pH (Figs. 3-5).

$$P = \sqrt{R^2 + B^2 + G^2} \tag{1}$$

$$\mathrm{r} = \frac{R}{P},\ \mathrm{b} = \frac{B}{P},\ \text{and}\ \ \mathrm{g} = \frac{G}{P} \tag{2}$$

$$\phi = atan2\,(g, r)\ \text{and}\ \ \theta = cos^{-1}(b) \tag{3}$$

The hue model is based on HSV (Hue–Saturation–Value) color space. Saturation (S) determines the intensity (or "colorfulness") of a color relative to its brightness: S = 0 indicates achromatic (gray) pixels with no color, while S = 1 represents fully saturated pure color. Value (V) is the brightness of the pixel, defined in [0, 1] where 0 indicates pure black and V = 1 means maximum brightness. For optical pH indicators, V correlates with local illumination intensity and overall raw frame contrast, while hue (H) represents the dominant color angle on the circular color wheel. H was calculated from RGB via eq.s 4 – 8 ( performed using HSV = rgb2hsv(RGB) function in MATLAB), where $h'$ is the unnormalized hue value on the RGB color wheel [0,6). $h'$ was normalized to $H = h'/6$, giving a unitless HSV hue coordinate $H \in [0,1]$ for calibration and pixel-wise pH mapping. Saturation-value-weighted circular mean was used to obtain a single representative hue value for each calibration image (see eq. 8b), where $w_i$ is the pixel weight. From calibration images, a monotonic interpolation function pH = f $(H)$ is fitted using a second-order polyfit, restricted to the calibrated pH range. This function is then evaluated pixel-wise on experimental time-series images to convert hue values to pH. Confidence intervals were estimated by bootstrap resampling (1000 resamples) of the calibration points (Fig. 2).

$$Value\ (V) = max\ (R, G, B) \tag{4}$$

$$Chroma\ (C) = V - \min(R, G, B) \tag{5}$$

$$Saturation(S) = \begin{cases} 0, & if\ V = 0 \\ \frac{C}{V} & V \neq 0 \end{cases} \tag{6}$$

$$h' = \begin{cases} 0, & if\ \ C = 0 \\ \left[\frac{G - B}{C}\ mod\ 6\right], & if\quad V = R \\ \left[\frac{B - R}{C} + 2\right], & if\quad V = G \\ \left[\frac{R - G}{C} + 4\right], & if\quad V = B. \end{cases} \tag{7}$$

$$H = \frac{h'}{6},\ \ H \in\ [0,1] \tag{8}$$

$$w_i = S_i V_i \tag{8a}$$

$$\dot{H} = \frac{1}{2\pi} atan2\left(\textstyle\sum_i w_i\, sin(2\pi H_i), \sum_i w_i\, cos(2\pi H_i)\right) \tag{8b}$$

From eqs. 1 to 7, the two models (hue and θ–ϕ) both discard brightness. The hue removes it implicitly through chroma and color ratios, while θ–ϕ removes it explicitly by eliminating the vector magnitude during the angular transformation. The key difference is that the hue retains only one coordinate, H, on the color wheel, while θ–ϕ retains two coordinates (θ and ϕ) after angular transformation. We therefore investigate how these transformations affect measurement reliability.

The free-fluid and porous-domain pH calibration and mapping followed the same hue and θ–ϕ quantifications described above; however, the porous-domain calibration and mapping were restricted to fluid-filled pore pixels using the binary segmented mask (see fig S2). Both routes used the same pore-domain mask, and an additional RGB-magnitude threshold of 0.02 was applied to exclude unstable low-intensity pixels (representing approximately 1.12% of the pore-space pixels). For each frame, pH was evaluated only within the segmented pore region, explicitly excluding the solid pellet.

Once the two pH calibration model curves were established from the calibration images (Figs. 2-5), experimental RGB images (acquired after $CO_2$ injection) were converted via the two different color-space representations to pH maps. For the hue quantification route, a reliability mask based on saturation and brightness thresholds ($S \geq 0.08$ and $0.10 \leq V \leq 0.98$) was applied to identify pixels with meaningful color information, and the calibrated monotonic function (pH$= f(H)$) was then applied pixel-wise, generating a full pH map for each frame (Figs. 10-11). For the θ–ϕ model, the coordinates θ and ϕ were extracted, a magnitude threshold was applied to remove low-intensity pixels, and the calibrated 2-D interpolant pH$=F(\theta, \phi)$ was applied to every valid pixel to obtain the θ–ϕ-based pH maps (Figs. 10-11). The spatial mean pH (averaged over the entire image as a function of time) for all domains and geometries was also extracted (Fig. 8). The pH maps were converted to total dissolved carbon concentration ($C_T$) using the PHREEQC speciation equilibrium solver (section 2.7.1).

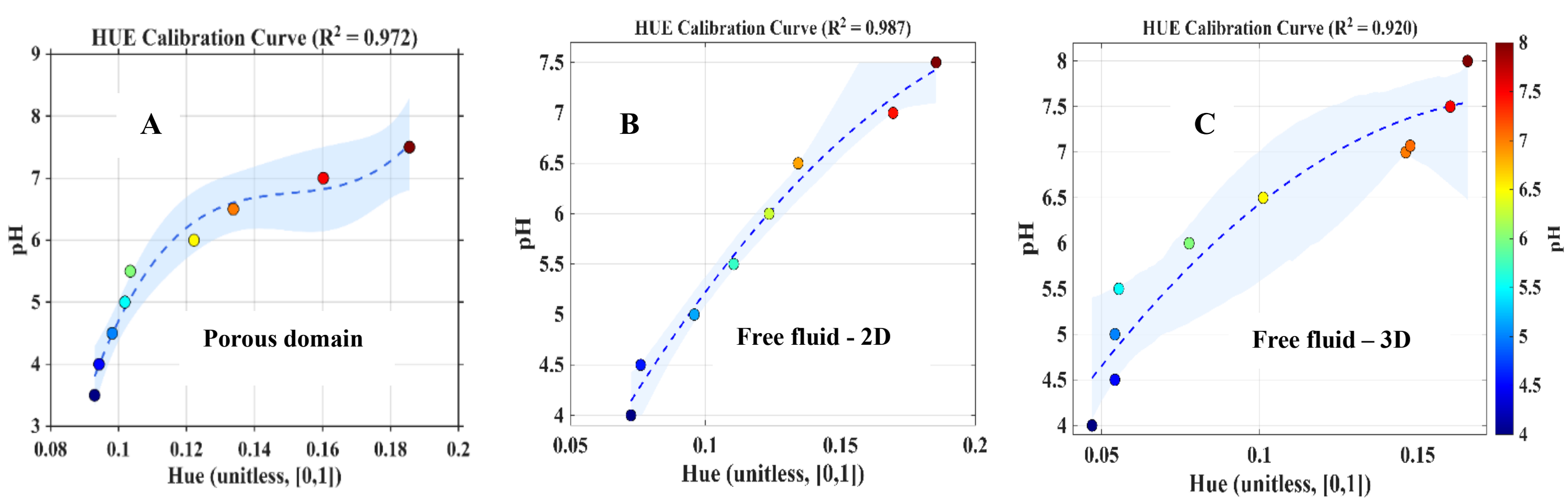

Fig. 2. Hue-based pH calibration curves show the relationship between normalized hue, H ∈ [0, 1], and measured pH for (a) porous medium, (b) 2D free-fluid, and (c) 3D free-fluid domains.

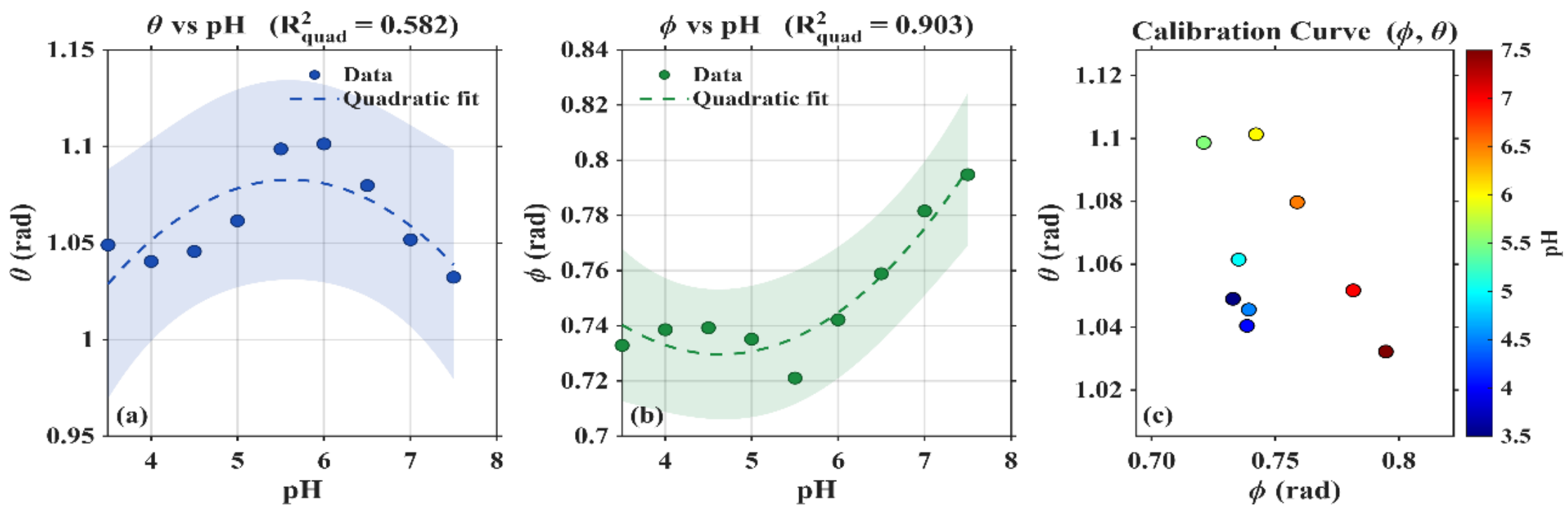

Fig. 3: θ–ϕ pH calibration curve for the porous domain.

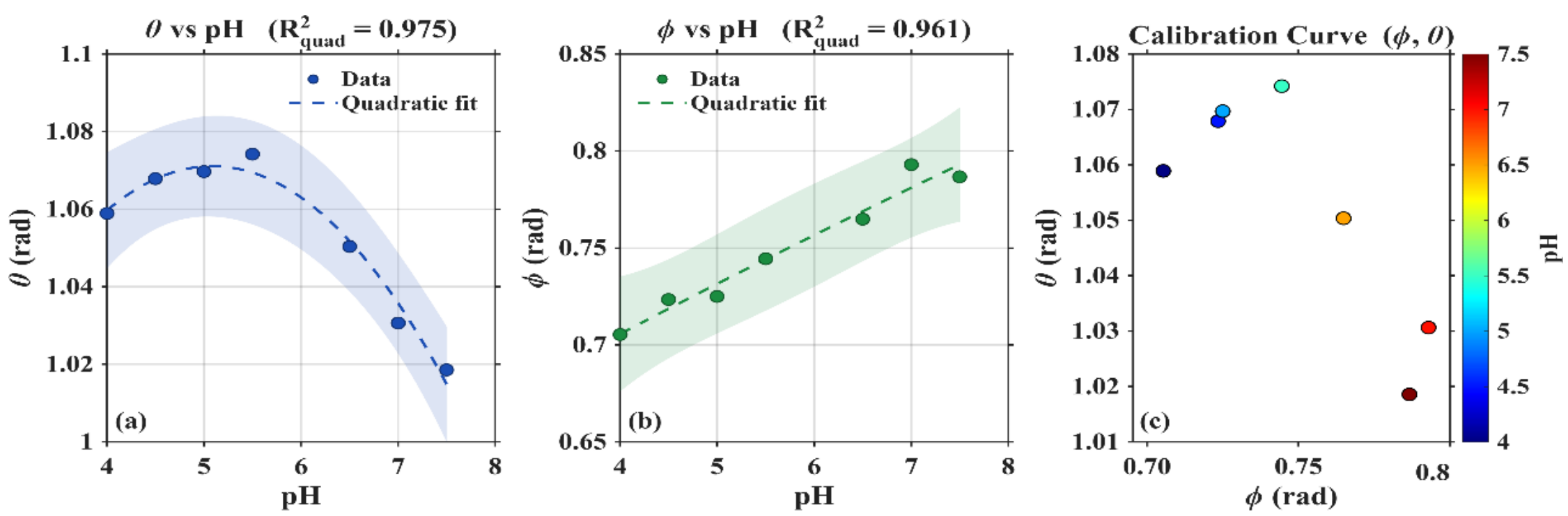

Fig 4: θ–ϕ pH calibration curve for the 2D free-fluid cell.

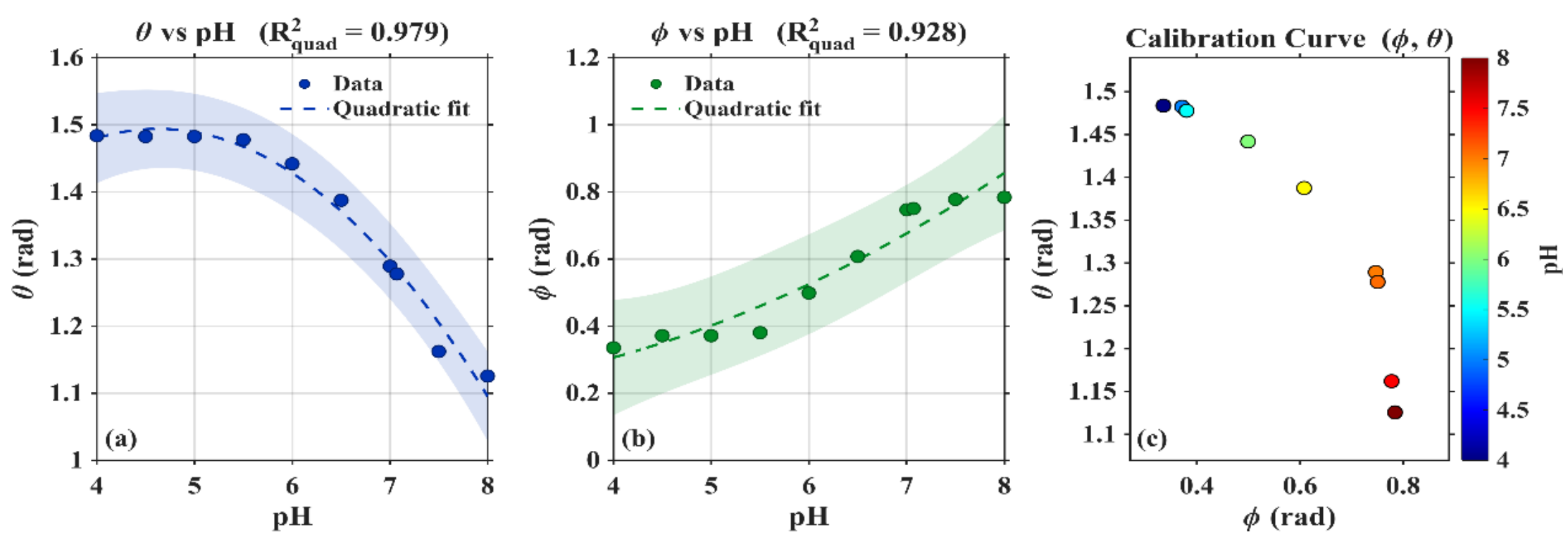

Fig 5: θ–ϕ pH calibration curve for the 3D free-fluid cell.

### 2.6 Measurement Reliability Test

Four constraint tests were designed to quantify the effects of experimental conditions and color model-dependent factors on spatial colorimetric pH and $C_T$ mapping. These tests included the effects of indicator mix concentration on pH calibration (see Supplementary Materials, Sec 3.0), pH calibration interpolation error and sensitivity, temporal and spatial stability, and propagation of measured pH noise into dissolved-carbon estimate. As noted above, the low concentration mix 1 was not viable for use in the porous medium domain, so subsequent tests only utilized mix 2. Together, the tests were used to determine whether each color model provided reliable mapping across different domains and geometries and to identify the conditions under which pH and $C_T$ measurements became less dependable.

### 2.6.1 pH Calibration Interpolation Error and Sensitivity Test

The calibration curve interpolation error and the susceptibility to noise amplification were tested to evaluate the domain and geometry-dependent performance of the two colorimetric models. For both models, the interpolation error was assessed using leave-one-out cross-validation (LOOCV). Each calibration point $i$ was iteratively removed, and the pH calibration function was refitted using the remaining $N-1$ calibration points. The omitted pH was predicted based on the measured H or θ and ϕ values. The prediction error (ei) for each omitted point in the models was calculated using Eq. 9, where $pH_i$ is the known pH of the omitted calibration frame and $p\widehat{H}_i$ is predicted by the calibration model built without that known point. The overall interpolation error was expressed as the LOOCV root mean square error (RMSE) and LOOCV mean absolute error (MAE) (eq.s 10-11). A low LOOCV RMSE indicates that the calibration is accurate at the calibration points.

The sensitivity to noise amplification ($S_H$) of the hue was evaluated from the slope of the fitted curve over the calibrated hue range, whereas sensitivity to noise of θ–ϕ curve ($S_{\theta\phi}$) was evaluated from the spatial gradient of the interpolated pH surface (pH= $F(\theta, \phi)$ in angular space (see Figs. 3-5). For hue, a grid was generated between the minimum and maximum calibration values, and the fitted pH values were evaluated across this grid, whereas in the case of spherical coordinates, a grid was also generated over the calibrated $\theta$ and ϕ ranges, and the interpolated pH surface was evaluated. $S_{\theta\phi}$ was then calculated from the gradient magnitude (see eq. 12), where pH = F(θ,ϕ). The 95th-percentile (P95) values of $S_{\theta\phi}$ and $S_H$ were calculated as P95 surface sensitivity; a larger value indicates that a small angular color error could propagate into larger pH noise. The P95 $of\ S_{\theta\phi}$ and $S_H$ provide a conservative indicator of the regions of the calibration surface most susceptible to color noise. The normalized hue $H \in [0,1]$represents one full hue-wheel rotation, equivalent to $2\pi$ radians, so the P95 of $S_H$ was converted from pH per normalized hue unit to pH/rad to be compared to the P95 $of\ S_{\theta\phi}$(Table 3 and 4). The P95 sensitivity results ($S_{\theta\phi}$ and $S_H$) should not be interpreted as pH error; it only describes how steep the calibration function is in color-coordinate space.

Furthermore, interpolation support in the θ–ϕ calibration space was evaluated by computing the number of support points, nearest-neighbor support spacing, and convex-hull area. The support points represent the angular calibration anchors used to construct the scattered interpolant. For each support point $j$, the nearest-neighbor spacing was computed using Eq. 14.

A convex ($A_{hull}$) was computed from the outer boundary of the support points in $(\theta, \varphi)$space, which represents the angular region over which interpolation is supported by calibration data. A larger convex hull area indicates broader calibration space coverage. Finally, simple coordinate-level $R^2$values were calculated for $\theta$ versus pH and $\phi$versus pH (Figs. 3-5).

$$ei = p\widehat{H}_i - pHi \quad (9)$$

$$MAE_{LOOCV} = \frac{1}{Nv}\sum_{i=1}^{Nv} |ei| \quad (10)$$

$$RMSE_{LOOCV} = \sqrt{\frac{1}{Nv}\sum_{i=n}^{Nv}(ei)^2} \quad (11)$$

$$S_{\theta,\phi} = \nabla pH = \sqrt{\left(\frac{dpH}{d\theta}\right)^2 + \left(\frac{dpH}{d\phi}\right)^2} \quad (12)$$

$$S_H = \left|\frac{d(pH)}{dH}\right| \quad (13)$$

$$d_j = \min_{k \neq J} \sqrt{\left(\theta_j - \theta_{k'}\right)^2 + \left(\phi_j - \phi_k\right)^2} \quad (14)$$

$$A_{hull} = \text{area of convex hull of } (\theta,\varphi) \text{ support points} \quad (15)$$

**2.6.2 Temporal and Spatial Stability Test**

The stability test was carried out by filling the flow cells of different domains and geometries with the indicator mix solution at a fixed pH (7.22) and imaging them for at least 11 hours to quantify the spatial and temporal noise of the indicator solution itself in absence of injected $CO_2$. The stability test was conducted at pH 7.22 because the standard probe-measured initial pH of the indicator solution used for $CO_2$-dissolution experiments was 7.22.

The test was evaluated at two levels: (1) on raw RGB frames, and (2) on their respective generated pH maps. This establishes the baseline noise in the raw frames as well as noise after the frames have been converted to pH maps. For the raw frames, the extracted RGB values were converted to H, θ, and ϕ, with no pH relation. The RGB brightness (I) and magnitude (r), as well as pixel-wise spatial and temporal variability of the corresponding H, θ, and ϕ, were estimated (Figs. S8 a and b).

In the case of the fixed pH maps, the spatial and temporal noise was quantified by computing these four metrics: pH bias, drift of slope per min, temporal standard deviation ($\sigma_{temporal}$ of mean pH), and median spatial standard deviation using eq.s 18-21, where $pH_{i,k}$ is the pH value of valid pixel i at frame k, $t_k$ is the time, M is number of frames, $N_k$ is number of valid pixels, $\overline{pH_k}$ is the mean pH of frame k, $\overline{pH}$ is the average pH across all analyzed frames, $pH_{true}$ is the probe-measured pH.

The pH bias accounts for the offset between the mean pH and the probe-measured pH. The temporal standard deviation measures frame-to-frame fluctuations in the domain-averaged pH. The median spatial std quantifies the pixel-to-pixel spatial variability of pH maps. Furthermore, to understand the spatial spread within each pH map, the 95th and 5th percentiles of pixels were evaluated, and the spread ($S_k = P95_k - P5_k$) over time, as well as the median for each data set, was calculated. Likewise, the P95 of the pixel temporal standard deviation was computed to identify the upper-bound temporal noise exhibited by the noisiest 5% of valid pixels (Table S4).

$$I = \frac{(R + G + B)}{3} \tag{16}$$

$$r = sqrt(R^2 + G^2 + B^2) \tag{17}$$

$$\text{Mean pH } (\overline{pH}) = \frac{1}{M}\sum_{k=1}^{M} \overline{pH_k} \tag{18}$$

$$Bias_{pH} = pH_{true} - \overline{pH} \tag{19}$$

$$\sigma_{pH}^{temporal} = \sqrt{\frac{1}{M-1} \sum_{k=1}^{M} (\overline{pH}_k - \overline{pH})^2} \tag{20}$$

$$\sigma_{pH,k}^{spatial} = \sqrt{\frac{1}{N_k - 1} \sum_{i=1}^{N_k} (pH_{i,k} - \overline{pH_k})^2} \tag{21}$$

**2.6.3 pH and $C_T$ Noise Propagation**

The pH uncertainty quantified from the fixed pH map stability test was propagated into the $CO_2$ dissolution pH maps using Monte Carlo (MC) noise propagation approach. The fixed-pH map noise, rather than the raw RGB frame noise, was used for MC propagation because it accounts for the cumulative noise across the full spatial colorimetric workflow (from RGB to pH calibration curves to pixel-wise pH mapping). For each dataset, the temporal and spatial pH standard deviations obtained from the stability analysis were used as the noise amplitudes, denoted as $\sigma_{pH,t}$and $\sigma_{pH,s}$, respectively. For each $CO_2$ dissolution pH map, Gaussian noise was independently added to every valid pixel using Eq. 23, where $pH_{i,j}^{raw}$is the original pH value at pixel $(i, j)$, $\sigma_{pH}$is either the temporal or spatial pH noise from the stability test, and $Z_{i,j} = N(0,1)$ is a normally distributed random variable (Figs. S9-S13).

The noisy pH values were restricted to the calibrated pH range of 3.5 –7.5 before conversion to total dissolved carbon. This procedure was repeated for 100 MC realizations for each pH map and noise type using MATLAB. Both the original (raw) and noisy pH maps were then converted to $C_T$ using the PHREEQC pH–$C_T$ lookup table (see Section 2.7.1 and eq. 22).

The absolute change in domain pH and the relative change in $C_T$ after the noise propagation were evaluated. For each MC realization, Gaussian noise was added pixel-wise to the original pH map to generate a perturbed pH map. Eq. 24 was then used to aggregate the pixel-wise difference between the original and perturbed pH fields over the valid mapping region, providing a map absolute change in domain mean pH ($\Delta pH$ ) (Fig 7a). In equations 23-24, $pH_i^{raw}$is the original pH value at pixel $i$, $pH_i^{noisy}$is the noise-perturbed pH value, $N$ is the number of valid pixels.

The relative change in domain mean $C_T$ (%$C_T$) was also computed to quantify the percentage change in domain-mean $C_T$ relative to the unperturbed $C_T$ map using eq. 25, where $C_T^{raw}$is the dissolved carbon map from the original pH map, while $C_T^{noisy}$is the dissolved carbon map after the measured pH noise has been propagated into the pH map (Fig 7b). Additionally, to capture the individual pixel local changes caused by the imposed noise, the absolute change in pH and $C_T$ was calculated at each valid pixel for each MC realization before averaging; the resulting mean pixelwise changes were reported in pH units and mol $L^{-1}$ respectively (Fig. S14)).

$$C_T = f_{\text{PHREEQC}}(pH) \tag{22}$$

$$pH_{i,j}^{noisy} = pH_{i,j}^{raw} + {}_{\sigma_{pH}}\ Z_{ij} \tag{23}$$

$$\Delta pH_{(domain\ mean\ )} \ = \ \frac{\left|\sum_{i=1}^{N}(pH_i^{noisy} - pH_i^{raw})\right|}{N} \tag{24}$$

$$\%C_T\ (domain\ mean) = 100\ \frac{|\sum_{i=1}^{N} CT_i^{noisy} -\ \sum_{i=1}^{N} CT_i^{raw}|}{(|\sum_{i=1}^{N} CT_i^{raw}|)} \tag{25}$$

## 2.7 pH and $C_T$ Measurement, Conversion, and Scaling

### 2.7.1 PHREEQC-based equilibrium chemistry and pH-to- Ct conversion

The PHREEQC equilibrium solver was used to establish the thermodynamic reference chemistry of the prepared indicator solution. To run the equilibrium simulation, the total alkalinity (TA) was first estimated by equilibrating the indicator solution with atmospheric $CO_2$ at 25 °C, using $pCO_2$ of 420ppm and adjusting the prescribed alkalinity until the simulated pH matched the measured indicator initial pH, ($pH_0 = 7.22$). The dissolved ionic composition of the indicator mix solution was defined from the ions of the indicator salts. Bromothymol blue ($C_{27}H_{27}Br_2NaO_5$) and thymol blue ($C_{27}H_{29}NaO_5S$) sodium salts constituted $Na^+$ = 1.286 x $10^{-4}$mol /L and $Cl^-$ =1.05 x $10^{-4}$mol /L from the chloride in methyl red ($C_{15}H_{15}ClN_3O_2$). This gave a total alkalinity (TA) of 1.207 x $10^{-4}$ mol /L as $HCO_3^-$. The TA was then used in a second PHREEQC simulation in which the same solution composition was equilibrated with $CO_2$ at a partial pressure of 1 atm. The resulting pH was taken as equilibrium pH ( $pH_{eq} = 4.14$), and the PHREEQC total dissolved inorganic carbon output C(4) = [ $CO_{2aq}$] + [$HCO_3^-$ ] + [$CO_3^{2-}$] was used as the saturated dissolved inorganic carbon concentration ($C_{sat\,=}$ 0.0348 mol/l). $pH_{eq}$ and $C_{sat}$ were used to normalize pH and carbon dissolution limits across the different domains and geometries. To convert all pH vs. time data from $CO_2$ dissolution experiments to $C_T$, a PHREEQC pH–$C_T$ lookup table (LUT) was generated over the calibration pH range of 3.5–8.00 with a pH interval of 0.01. In the LUT, pH was prescribed over the grid using the same TA, $Na^{+,}$ and $Cl^-$, producing a PHREEQC-calculated value of $C_T$.

To avoid nonphysical pH and $C_T$ estimates caused by pixel-level color noise, pH and $C_T$ values were capped at the PHREEQC-predicted equilibrium condition, ( $pH_{eq}$ and $C_{sat}$ ).

### 2.7.2 Dissolved Inorganic Carbon Measurement

Two dissolved total inorganic carbon concentrations were estimated: the integrated $C_T$ ($N_{CT}$) and the mean $C_T$ ( $\overline{C_T}$).

The $\overline{C_T}$ values compare average concentrations (a proxy for dissolution response) across domains independent of the volume, whereas the $N_{CT}$ was used to estimate the total dissolved inorganic carbon within each domain volume. $N_{CT}$ was estimated using eq. 26, where $C_{T,i}$ is the total dissolved inorganic carbon concentration in the domain (mol/L). V is the physical fluid volume represented by one image pixel. $d_x$ and $d_y$ are the physical pixel dimensions, and b is the flow-cell gap. The $N_{CT}$ for the 3D data set was classified as "apparent" because the acquired image is a 2D projection of the 3D geometry.

For all analyses of mean pH and $C_T$, the diffusive stable layer at the gas-water interface (see fig S20) was excluded prior to conversion to pH maps, as the high local $C_T$ in that region reflects interfacial accumulation rather than the transport within the convective bulk region that this analysis aims to quantify. Spatial pH maps were converted into a $C_T$, and the mean $C_T$ was calculated. Mean pH and $C_T$ was plotted against the physical time (fig 8). To further compare the convective dissolution curve shape and gradient across the different domains, mean pH and $C_T$ were scaled using the convective dimensionless time (t*) and diffusive dimensionless time ($t^*_{diff}$) via eq.26b where the characteristic time ($t_c$) is the time required for buoyancy-driven flow to move through the domain height H at the characteristic convective velocity ($U_c$), t = physical time, D = molecular diffusivity of dissolved $CO_2$, μ = dynamic viscosity, Δρ = density difference, and $\phi_p$ is porosity.

$$N_{CT} = \sum_{i=1}^{n} C_{T,i}\,(t)V \quad (26)$$

$$V = d_x\, d_y b \quad (26a)$$

$$t^* = \frac{t}{tc},\ t_c = \frac{\phi H}{U_c},\ U_c = \frac{k\Delta\rho g}{\mu},\ \ t^*_{diff} = \frac{tD}{H^2} \quad (26b)$$

### 2.7.3 pH and $C_T$ Evolution Curves Shape Agreement and Gradient Analysis

The maximum temporal response across all domains and geometries was defined as the largest absolute gradient of mean pH or $C_T$ with respect to convective dimensionless-time curves ( max $|\, dpH/dt^* \,|$ and max $| dC_T/dt^* |$) to compare the peak responses of the colorimetric models to the $CO_2$ dissolution. The shape of temporal curves (mean pH and $C_T$ vs t*) were also evaluated to determine whether hue and $\theta - \phi$ quantification methods captured similar temporal trends of $CO_2$ dissolution within the same geometry, independent of differences in quantified absolute pH, $C_T$, or experiment duration. For each paired dataset, the $t^*$axis was normalized to a common progression coordinate s, and the pH and $C_T$ curves were then normalized to a unit range ($Y_{norm}$), where Y represents either mean pH or mean $C_T$. This normalization removed differences in absolute magnitude and allowed the comparison to focus on curve shape. The normalized hue and $\theta - \phi$ curves were interpolated onto a common $s$-axis using pchip interpolation, and agreement was quantified over the overlapping data range only. The difference between paired normalized curves was computed using eq. 29 and summarized using normalized RMSE. Low RMSE means hue and $\theta - \phi$ share similar temporal curve shape, while larger values indicate method-dependent differences in the resolved dissolution trend.

$$s = \frac{t^* - t^*_{min}}{t^*_{max} - t^*_{min}} \quad (27)$$

$$Y_{norm(s)} = \frac{Y(s) - Y_{min}}{Y_{max} - Y_{min}} \quad (28)$$

$$\Delta Y_{norm(s)} = Y_{nrom,Hue}(S) \; - \; (Y_{nrom,\theta-\varphi}(S)) \quad (29)$$

$$RMSE_{shape} = \sqrt{\frac{1}{N}\sum_{j=1}^{N}}\left[\Delta Y(s_j)\right]^2 \quad (29a)$$

**2.7.4 pH and $C_T$ Equilibrium Progress Scaling**

To compare how the dissolution processes in each of the domains and geometries progresses towards the established equilibrium condition ($pH_{eq}$ = 4.14 and $C_{sat}$ = 0.0348 mol/l), the mean pH and $C_T$ curves were converted to normalized progress variables, $\xi_{pH}$ and $\xi_{C_T}$ using eq. 30-31. $\xi_{pH}$ and $\xi_{C_T}$ were plotted against t* and $t^*_{diff}$. Subscript 0 indicates the initial pH or $C_T$ values of each dataset, and $pH_{eq}^{PHREEQC}$ and $C_{T,sat}^{PHREEQC}$ are the equilibrium values predicted by PHREEQC.

$$\xi_{CT}(t) = \frac{C_T(t) - C_{T0}}{C_{T\,sat}^{PHREEQC} - C_{T0}} \quad (30)$$

$$\xi_{pH}(t) = \frac{pH_o - pH(t)}{pH_o - pH_{eq}^{PHREEQC}} \quad (31)$$

# 3 Results and Discussion

## 3.1 Measurement Reliability

### 3.1.1 pH Calibration Interpolation Error and Sensitivity Test

The susceptibility of hue and θ–ϕ pH calibration curves to noise amplification and interpolation error was tested to evaluate the domain and geometry dependent performance and measurement reliability of the color models. In the free-fluid domain pH calibration, the 2D hue model showed the lowest cross-validated calibration error, with LOOCV-RMSE = 0.20 pH and P95 sensitivity = 6.76 pH/rad, followed by the 3D hue (Table 3 and 4). In contrast, the 2D θ–ϕ calibration had a higher LOOCV-RMSE of 0.39 pH and a high P95 surface sensitivity of 87.81 pH/rad as compared to the 2D hue and 3D θ–ϕ. Although θ and ϕ strongly correlated with pH in the 2D θ–ϕ calibration, the combined θ–ϕ interpolation surface (Fig 4C) was sensitive. The 3D θ–ϕ calibration was better conditioned; the coordinates also correlated well with pH, it occupied a broader angular calibration space, with a convex hull area of 0.10 rad², and had lower P95 sensitivity than the 2D θ–ϕ case (see Table 3). This indicates that θ–φ reliability improves when calibration points occupy a broad, well-ordered angular space rather than a compressed or poorly separated region.

In the porous domain**,** hue calibration was more stable and better conditioned than θ–ϕ. Both approaches yielded similar LOOCV-RMSE and LOOCV-MAE values, but the hue-based P95 sensitivity was 13.8 pH/rad, 33X lower than the θ–ϕ sensitivity of 465.6 pH/rad (Tables 3 and 4). This demonstrates that t reliability cannot be assessed from curves cross-validation error alone: in the porous domain, θ–ϕ produced a slightly lower LOOCV-RMSE than hue, but its P95 calibration sensitivity is approximately 33X higher than hue, indicating a much greater potential for small color-coordinate perturbations to be amplified into pH uncertainty. In addition, the porous θ–ϕ pH calibration curve behaved completely differently from the 2D θ–ϕ free fluid dataset: in the lower pH range (3.5 to 5.5), which is most relevant to $CO_2$ dissolution, calibration points were not properly ordered in the angular space, and separability was not strong (Fig 3). The weak performance of θ–ϕ in the porous domain is due to its bivariate angular transformation of RGB in space: unlike the scalar angular transformation of hue, θ–ϕ depends on a two-coordinate surface, $pH = f(\theta, \phi)$. The performance and reliability depend on a smooth correlation between these coordinates and pH, as well as the angular space they occupy, as confirmed by the free-fluid domain results. The unstable responses of θ and ϕ resulted in a more highly sensitive calibration curve surface than hue. Due to these constraints, the reliability of pH measurement using the θ–ϕ model in the porous domain was compromised, and the porous domain was quantified only via the hue method.

Table 3: θ–φ pH calibration interpolation error and surface-sensitivity metric

| Method | LOOCV RMSE (pH) | LOOCV MAE (pH) | H min | H max | P95 sensitivity (pH/rad) |
|---|---|---|---|---|---|
| Porous – hue | 0.73 | 0.55 | 0.093 | 0.186 | 13.80 |
| FF-2D – hue | 0.20 | 0.17 | 0.072 | 0.185 | 6.76 |
| FF- 3D – hue | 0.52 | 0.44 | 0.047 | 0.166 | 6.80 |

Table 4: Hue pH calibration interpolation error and sensitivity metrics

| Method | Support Points | LOOCV RMSE (pH) | LOOCV MAE (pH) | P95 sensitivity (pH/rad) | Convex Hull area ($rad^2$) |
|---|---|---|---|---|---|
| Porous – θ–ϕ | 12,403 | 0.63 | 0.52 | 465.62 | 0.02 |
| FF-2D – θ–ϕ | 13,208 | 0.39 | 0.37 | 87.81 | 0.01 |
| FF-3D – θ–ϕ | 16,358 | 0.51 | 0.46 | 28.56 | 0.10 |

### 3.1.2 Temporal and Spatial Stability Test

The temporal and spatial stability of raw fixed-pH RGB frames and their respective fixed-pH maps was quantified to establish the baseline noise of the indicator mix solution in the absence of $CO_2$ injection. The raw RGB frame stability test results show that the 3D free-fluid dataset had the most stable raw color fields, as indicated by lower standard deviations in brightness and RGB magnitudes, followed by the 2D and then the porous domains (Fig. S8a).

The post-conversion hue values exhibited slightly higher baseline spatial variability than the θ-ϕ values across the dataset; the spatial noise of 2D FF θ and ϕ values reported (θ = 0.007 and ϕ = 0.005 in pH unit) herein are similar to those recorded by Birggison et al. (2025) for a 2D Hele-Shaw cell (Fig. S8b).

Across all domains, and geometries (Fixed pH maps), the temporal noise was very small ($\sigma_{pH}$ <0.01) compared to spatial noise (median spatial $\sigma_{pH}$ = 0.037-0.150) (Fig. 6 a and b), indicating that the dominant noise contribution is due to spatial map heterogeneity rather than frame-to-frame drift. The 3D θ–ϕ recorded the lowest spatial noise (median spatial $\sigma_{pH}$ = 0.035), followed by the 3D and 2D hue. The porous hue showed a lower spatial noise than the 2D θ–ϕ (Fig 6b), and the corresponding P95–P5 spread confirmed the same spatial trend (table S4). The 2D θ–ϕ recorded the highest spatial noise (median spatial $\sigma_{pH}$ = 0.15), which we attribute to the highly sensitive 2D θ–ϕ calibration surface (P95 surface sensitivity).

This highlights the fact that raw fixed pH RGB frames can have relatively low spatial noise in angular space, a sensitive calibration surface can amplify these small variations into high pH noise. This also confirms the need to quantify baseline noise directly from pH maps rather than relying only on raw RGB frames: raw-frame stability metrics do not fully capture the noise in final pH maps even when they are propagated into pH and $C_T$ errors. The porous hue case was highly stable in time but retained moderate spatial variability, indicating that porous-domain mapping is feasible but remains hindered by pore-scale optical effects.

The quantified noise level across the entire dataset indicates that fixed pH indicator images exhibit spatial and temporal heterogeneity even when the chemistry is uniform. To resolve pH and $C_T$ during convective dissolution using colorimetric methods, these metrics must be quantified to know the uncertainties. For the experimental conditions investigated herein, the reliability of pH measurement is most impacted by spatial pH noise, ranging from 0.037- 0.150 pH units depending on the specific experimental domain.

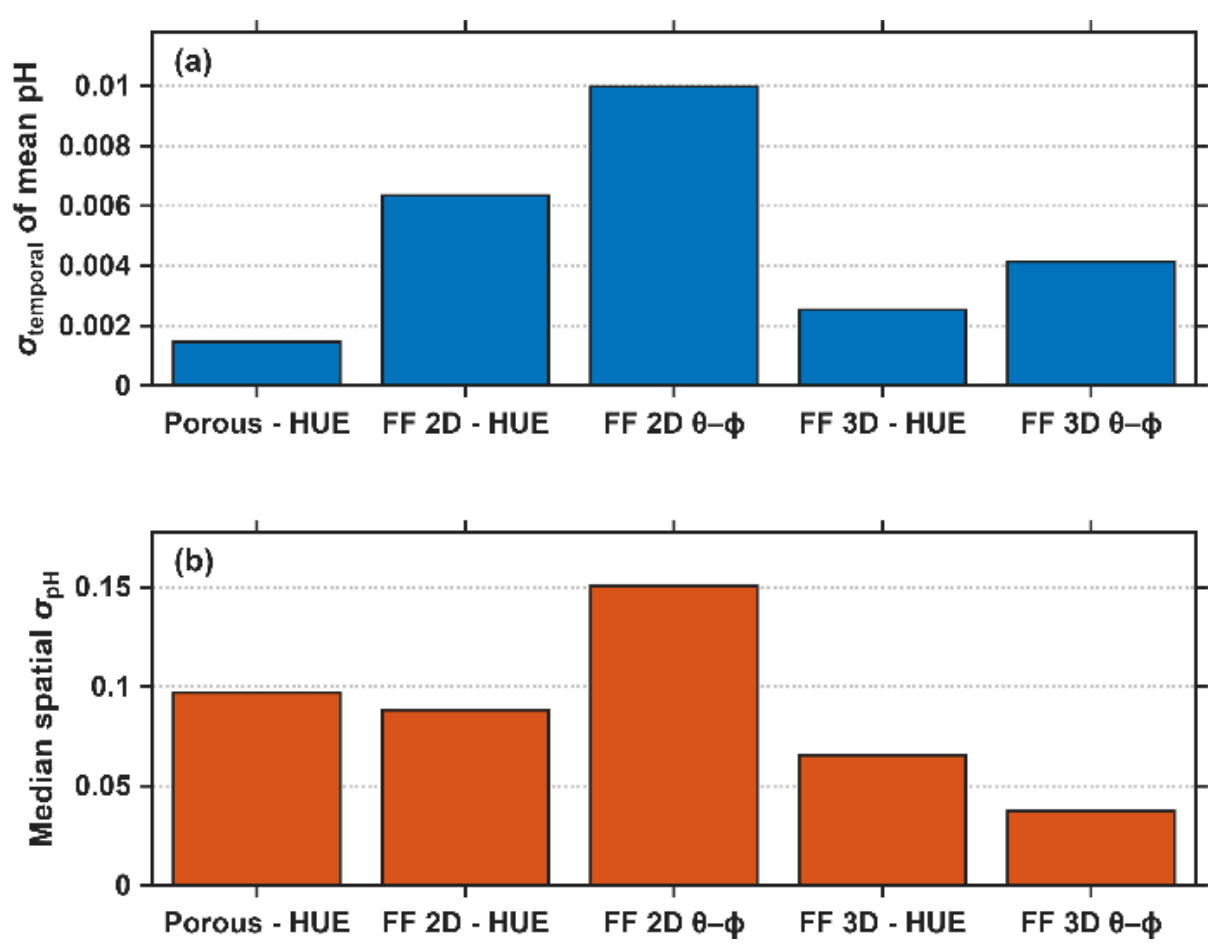

Fig 6: Fixed pH map stability test (a) temporal noise and (b) spatial noise.

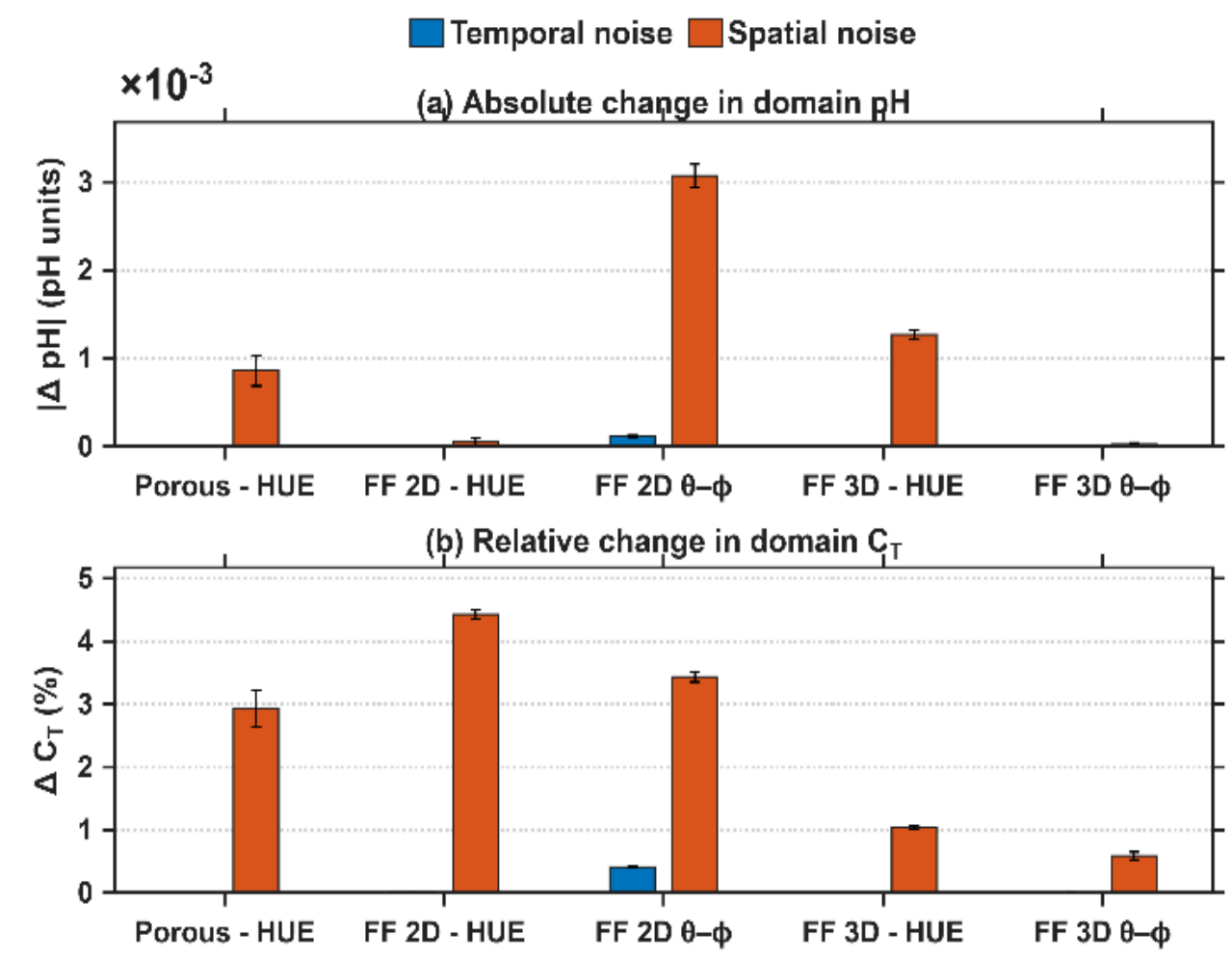

Fig 7: Domain-level changes in pH and $C_t$ after noise propagation. (a) Absolute change in domain-mean pH and (b) relative change in domain-mean $C_t$.

### 3.1.3 pH and $C_T$ noise propagation

The quantified baseline noise was propagated into $CO_2$ dissolution pH maps using the Monte Carlo (MC) noise-propagation approach. Across all dataset, temporal noise produced small domain-level changes (≤0.00012 pH units and ≤0.41% in $C_t$), whereas spatial noise produced pH changes of up to 0.0031 units and $C_t$ changes of 0.6 – 4.4% (Fig. 7). FF 2D θ–φ had the largest domain-mean pH change, while FF 2D hue had the largest relative $C_t$ change. So, a slight change in domain-mean pH does not necessarily imply a small change in derived $C_t$. Moreover, when local pixel changes were examined, the change in mean pH and $C_T$ were high (0.0298 – 0.115 pH unit and $6.2 \times 10^{-6}$ – $864 \times 10^{-6}$ $\Delta C_T$).

FF 2Dθ-ϕ showed the largest local changes, reaching 0.115 pH units and $864 \times 10^{-6}$ mol $L^{-1}$ in $C_t$ (about 7% change in mean $C_T$) (see fig S14), indicating that domain averaging can conceal local sensitivity to noise. Therefore, the reliability of dissolved-carbon mapping must be evaluated at the domain level and the pixel wise level. Also, the reliability of pH and $C_T$ measurement (Figs. 8 and 9) should be interpreted with the spatial domain level and local pixels uncertainties quantified for the different experimental domain and geometries.

## 3.2 pH and $C_T$ Measurement, Conversion, and Scaling

### 3.2.1 pH Spatial Maps

Spatial pH maps resolved by the hue and θ–ϕ models consistently captured the evolution of convective $CO_2$ dissolution across all domains and geometries (Figs. 10-11). In the free-fluid systems, both methods captured coherent downward-propagating plumes initiated from the top after the formation of a stable diffusive layer between the gas and liquid interfaces, with comparable plume positions and temporal progression. The θ–ϕ model generally produced sharper plume boundaries and stronger gradient contrast in 2D and 3D, while the hue model yielded smoother transitions and more spatially uniform background fields (Figs. 10 and 11). These differences arise from how each method retains color coordinates in space after transformation: because the hue model compresses chromatic information into a single coordinate, it produces smoother spatial transitions and reduced pixel-scale variability, particularly in regions of weak gradients such as the background fluid. (Figs. 10 and 11). In contrast, the θ–ϕ transformation preserves additional directional color information through its second coordinate, enabling stronger sensitivity to subtle color changes and sharper representation of plume or finger boundaries, but at the cost of higher sensitivity to noise.

This effect was most evident in the early stages of dissolution across all geometries (Figs. 10 and 11). Despite these visual differences, both methods identified the same physical plume geometry and evolution, indicating consistent detection of the underlying dissolution process.

Key differences emerge when comparing the convective fingers in the 2D and 3D free-fluid domains. In 2D, the developed finger remained relatively narrow, sharper and vertically coherent, with limited lateral spreading and gradual gradient transitions. This is attributed to the viscous force in the narrow channel: density driven flow develops slowly, mixing remains weak, and the finger boundary is sharper; so, hue and θ–ϕ methods produce relatively consistent pH contours. In the 3D system, the increased channel depth promotes convective motion and produces a broader mixing zone. As a result, plume structure in the 3D geometry appeared thicker and more irregular than in 2D, regardless of the colorimetric method used. Under these stronger convective conditions, differences between hue and θ–ϕ became more pronounced.

The porous domain maps exhibited highly localized dissolution pathways controlled by pore connectivity and tortuosity, as expected of a porous medium (Fig. 11). The hue model resolved the continuous finger-like penetration through connected pores; spatial variability persisted throughout the experiment. These patterns arise because fluid motion in porous media is constrained by solid grains, which fragment flow into discrete channels and limit large-scale mixing. Neighboring pores can experience different dissolution histories even at late times, producing persistent heterogeneity in both spatial maps and mean pH trends.

### 3.2.2 Dissolved Inorganic Carbon Measurement

The mean pH and $C_T$ extracted over physical time show that the 3D system achieved the lowest pH (4.79 – 4.71) and the highest $C_T$ response ($9.95 \times 10^{-3}$ – $6.6 \times 10^{-3}$ mol/l) for these experiment durations. The porous hue was intermediate, exhibiting a lower pH and higher $C_T$ response at the end of the experiment than the 2D free-fluid dataset. The ϕ–θ and hue models showed the same dissolution response trend for the free-fluid datasets; ϕ–θ produced the lowest pH and the highest $C_T$ response in both 2D and 3D (Fig.8 a & b). In terms of the colorimetric model response to post-injection effects, the 2D hue and 2D ϕ–θ showed the same stable late-time pH and $C_T$ plateaus, although ϕ–θ produced a significantly higher $C_T$, whereas the 3D ϕ–θ and 3D hue pH and $C_T$ plateaued shortly between 4000s and 5500s, followed by a pH rebound. The rebound is due to gas–liquid carbonate re-equilibration after $CO_2$ supply termination.

The Integrated $C_T$ showed a similar dissolution trend in the mean $C_T$ except for the porous hue. The porous hue had a higher mean $C_T$ than the 2D dataset; however, it's integrated $C_T$ was 3.3 x $10^{-6}$ mol, below the 2D ϕ–θ (Fig. 9). This apparent discrepancy is only because the Teflon packing reduces the total pore-water volume available for carbon accumulation.

The mean pH and $C_T$ were also scaled via the convective ($t^*$) and diffusive ($t^*_{diff}$) dimensionless time scales to enable cross-domain comparison. The scaling shows that each domain occupied its own convective timescale, with the same dissolution trend observed for mean pH and $C_T$ versus physical time (see Figs. S15 and S16).

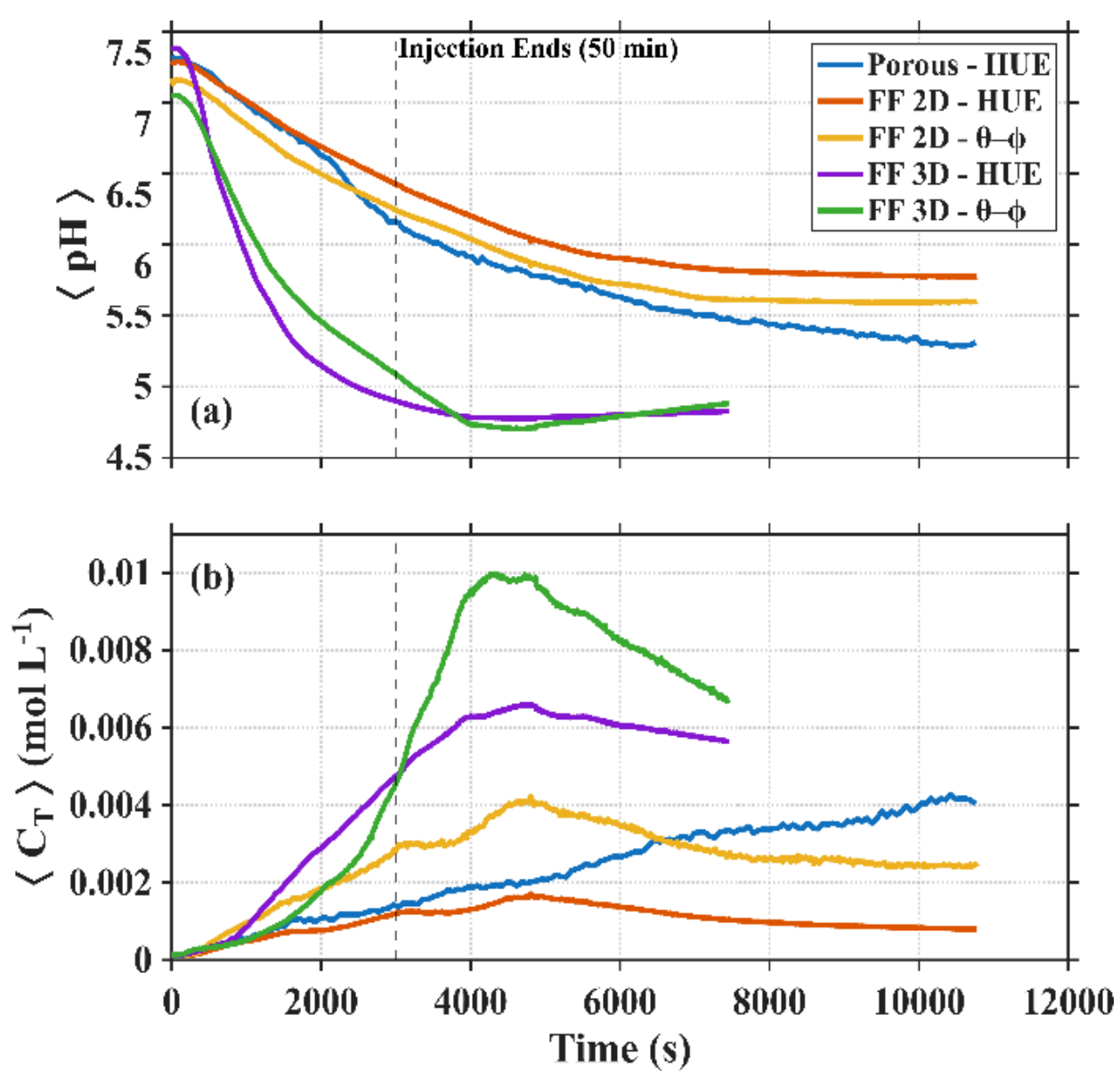

Fig 8: Mean pH and $C_T$ vs time.

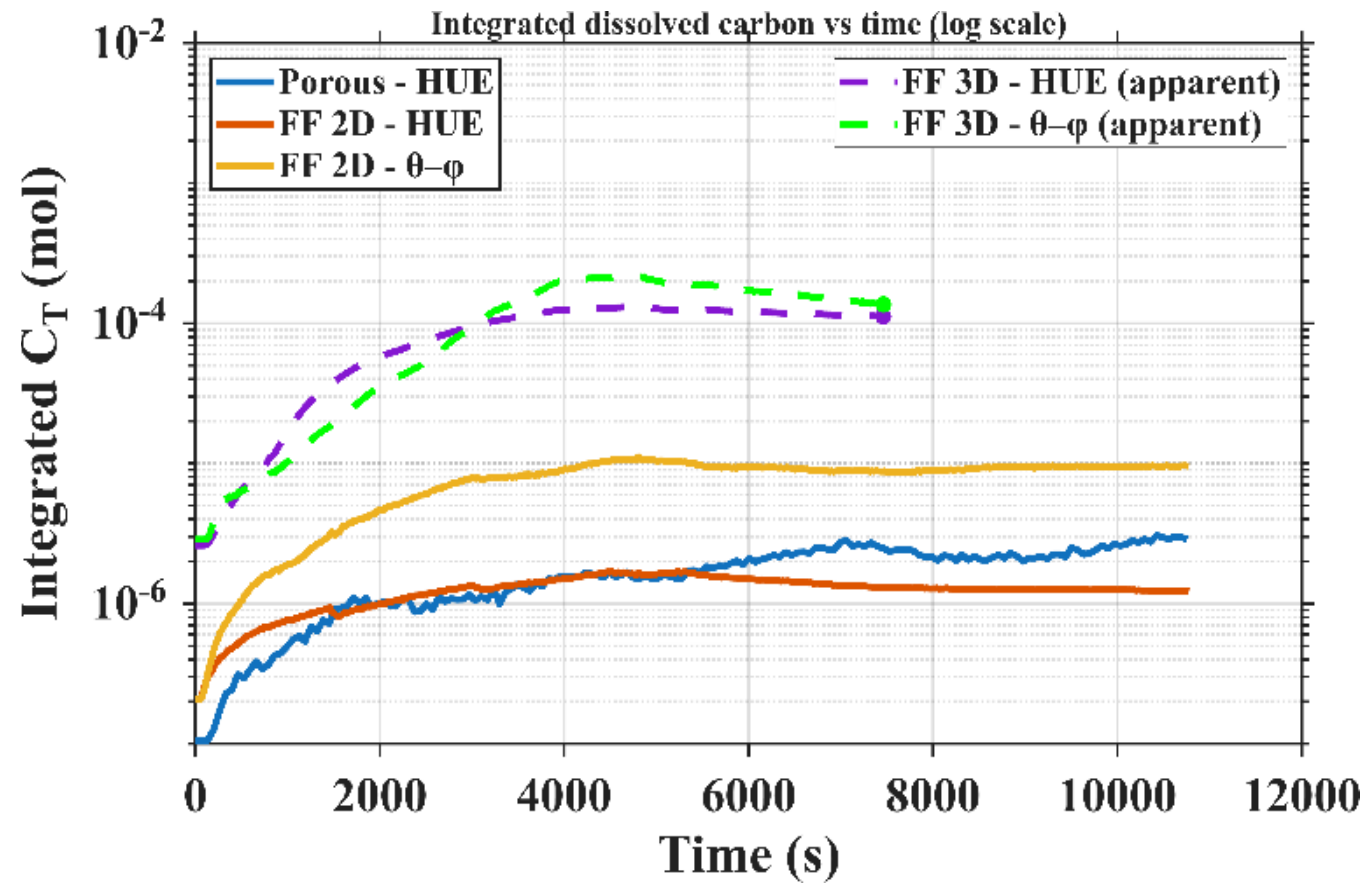

Fig 9: Integrated $C_T$ vs time for the different experimental domains. Dashed lines indicate "apparent" integrated $C_T$, as for the 3D cell the acquired image is a 2D projection of the 3D geometry.

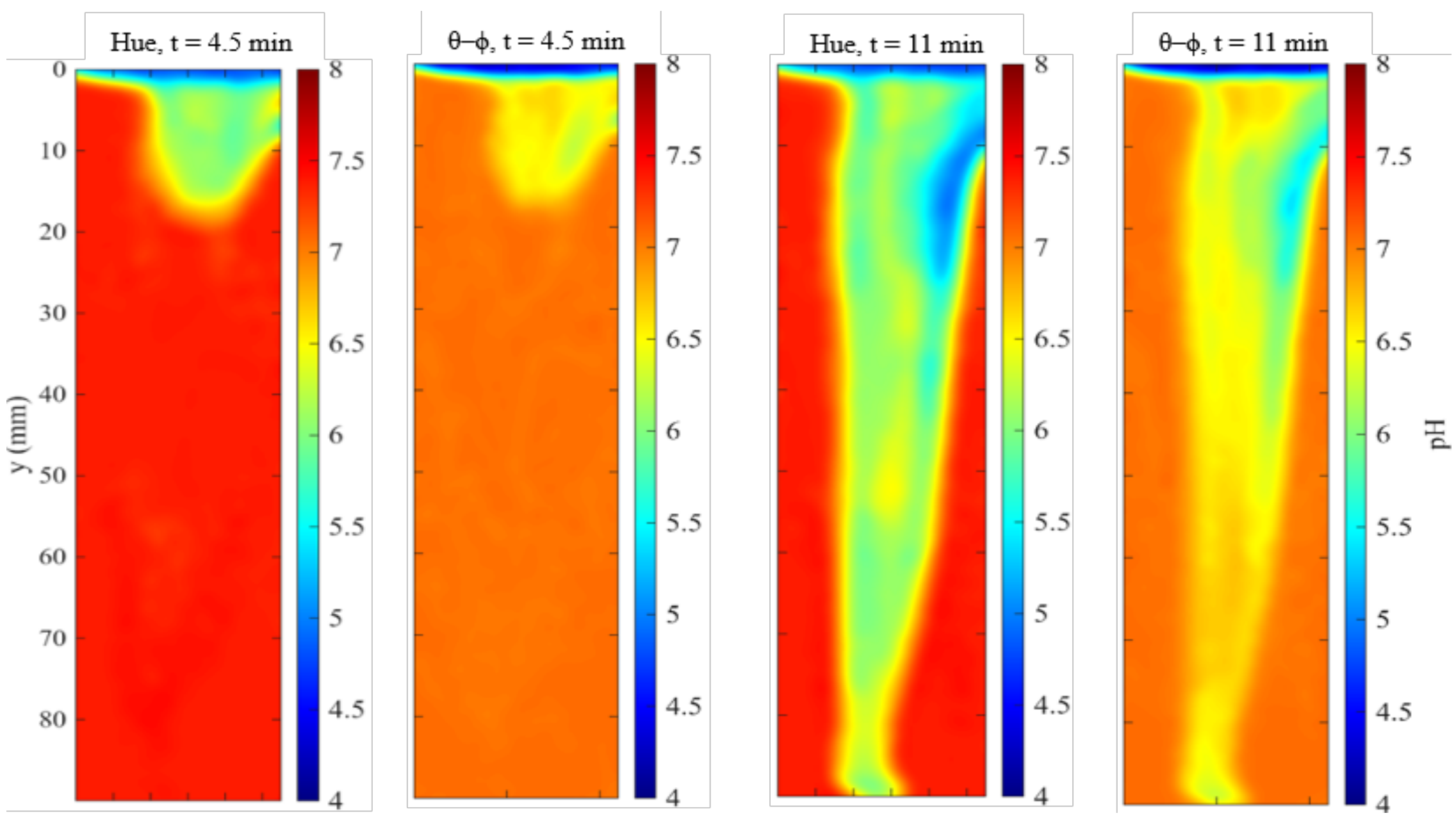

Fig 10: 3D hue and 3D θ–ϕ spatial

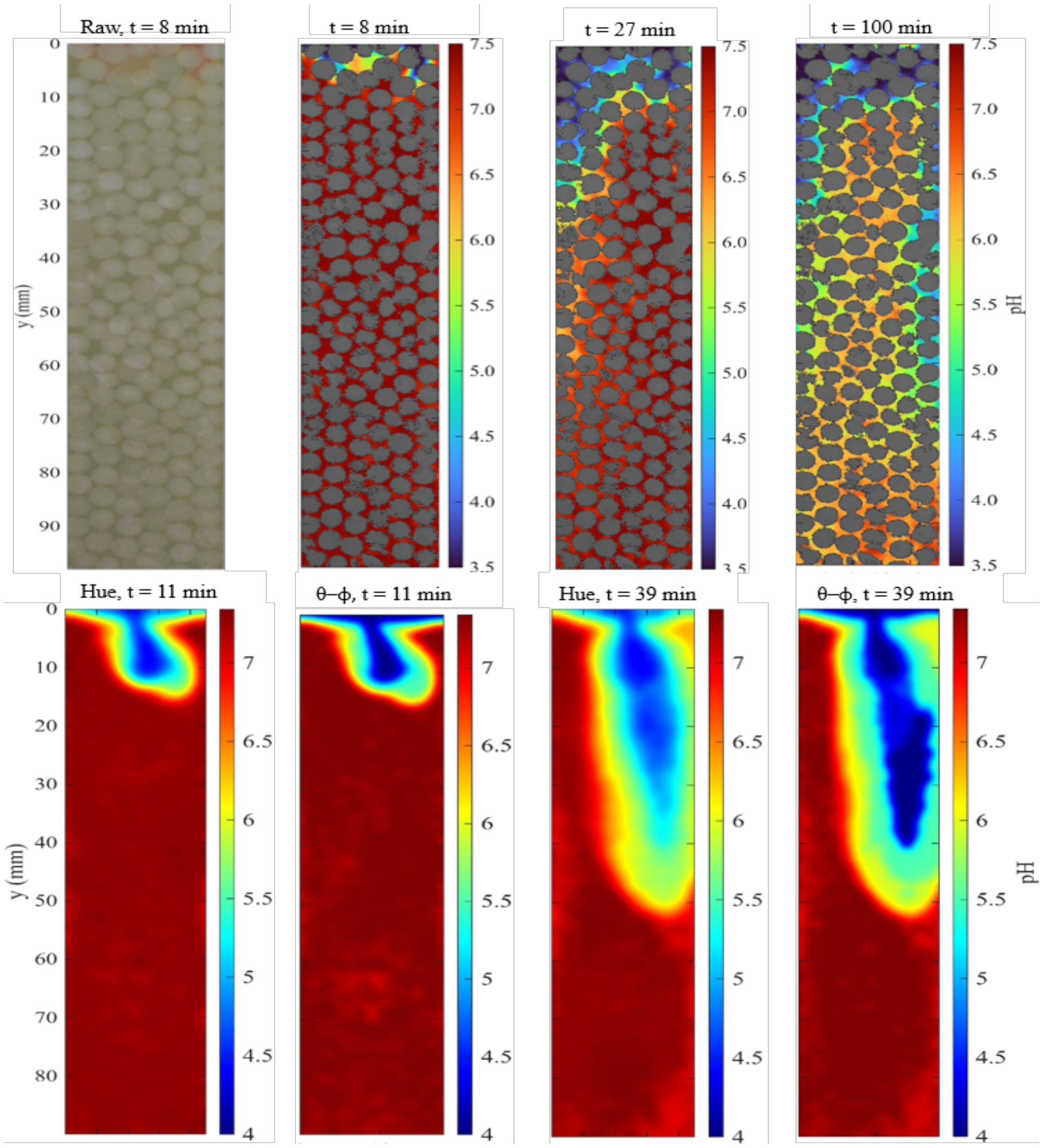

Fig 11: Spatial pH maps: porous-hue domain (upper panel), and FF 2D hue and θ–φ (lower panel), showing the spatial evolution of $CO_2$-rich convective fingers. The 2D FF maps compare colorimetric model-dependent differences in plume sharpness, gradient contrast, and background smoothness.

### 3.2.3 pH and $C_T$ Evolution Curves Shape Agreement and Gradient Metrics

The agreement between the shapes of the mean pH and $C_T$ vs time curves extracted using the hue and θ–ϕ models in 2D and 3D cells was evaluated based on the normalized progression of pH and $C_T$ (Fig 12). In 2D FF, hue and θ–ϕ shape agreement was very strong in the normalized pH response, with nearly overlapping trajectories across the full progression range. The normalized $C_T$ curves followed the same overall trend, rising to a mid-progression peak before declining, although it showed a modest late-time separation, with θ–ϕ maintaining slightly higher $C_T$ than hue (Fig 12). The normalized RMSE confirms the same trend: pH RMSE close to zero and $C_T$ RMSE = 0.061. (Figs S17). In 3D, the shape agreement between the two-color models was not strong for either pH or $C_T$ (Fig 12). The fact that the shape agreement was stronger in 2D than in 3D demonstrates that the wider optical path increases cell sensitivity to optical averaging and color-model-dependent conversion effects, amplifying differences in calculated $C_T$. The maximum gradients (max (|dpH/dt*|)) of pH and $C_T$ evolution curves were quantified to obtain the maximum temporal response of the data set. The results (Table S5) indicates that the temporal evolution of $CO_2$ dissolution extracted via hue and θ–ϕ in the free-fluid domains was internally comparable within each geometry but not identical across geometries. The color model can affect the apparent peak response within a geometry; however, the much larger differences arise from the occupied convective time scale.

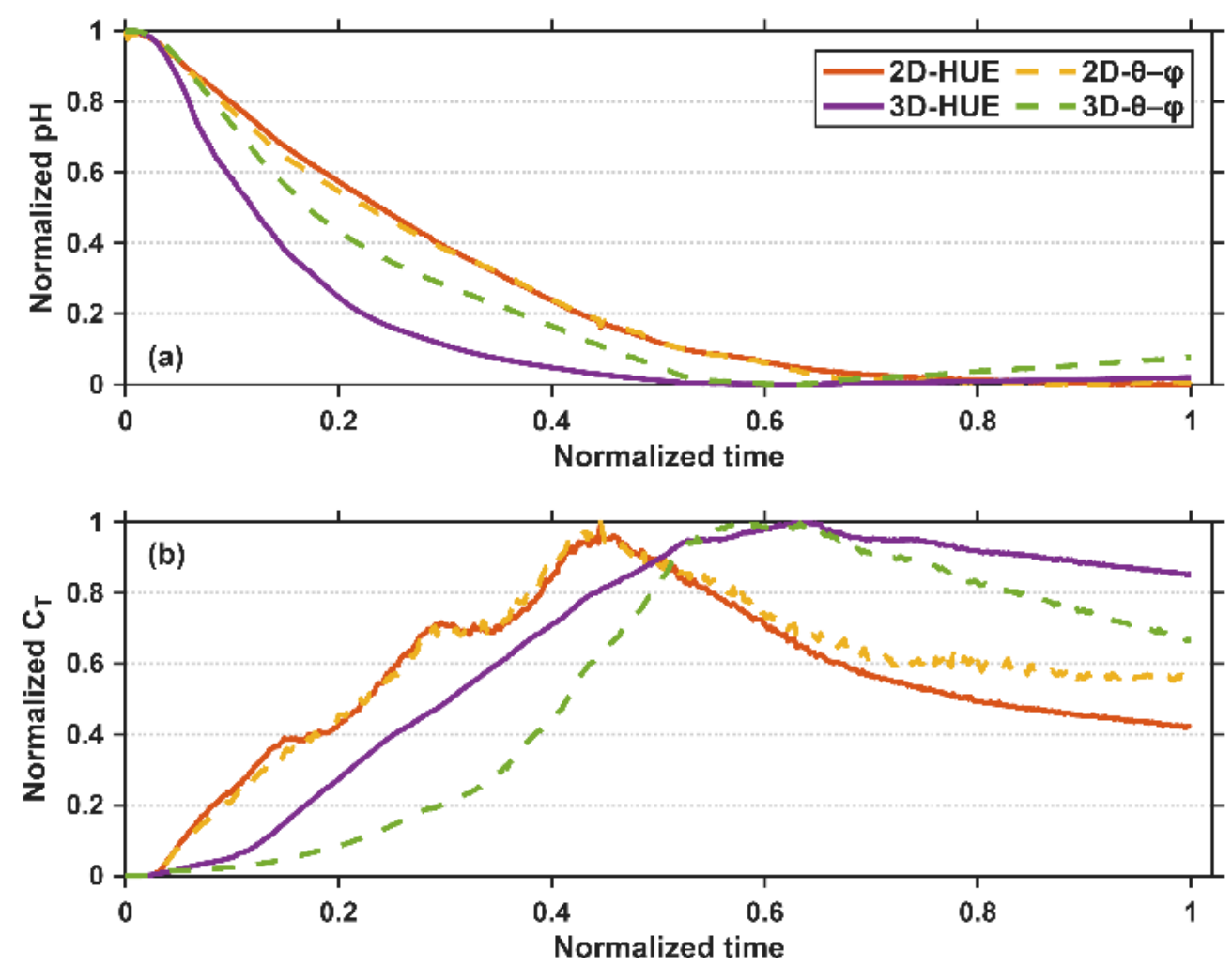

Fig 12: Hue and θ–φ pH and $C_T$ evolution curves shape agreement in FF 2D and FF 3D.

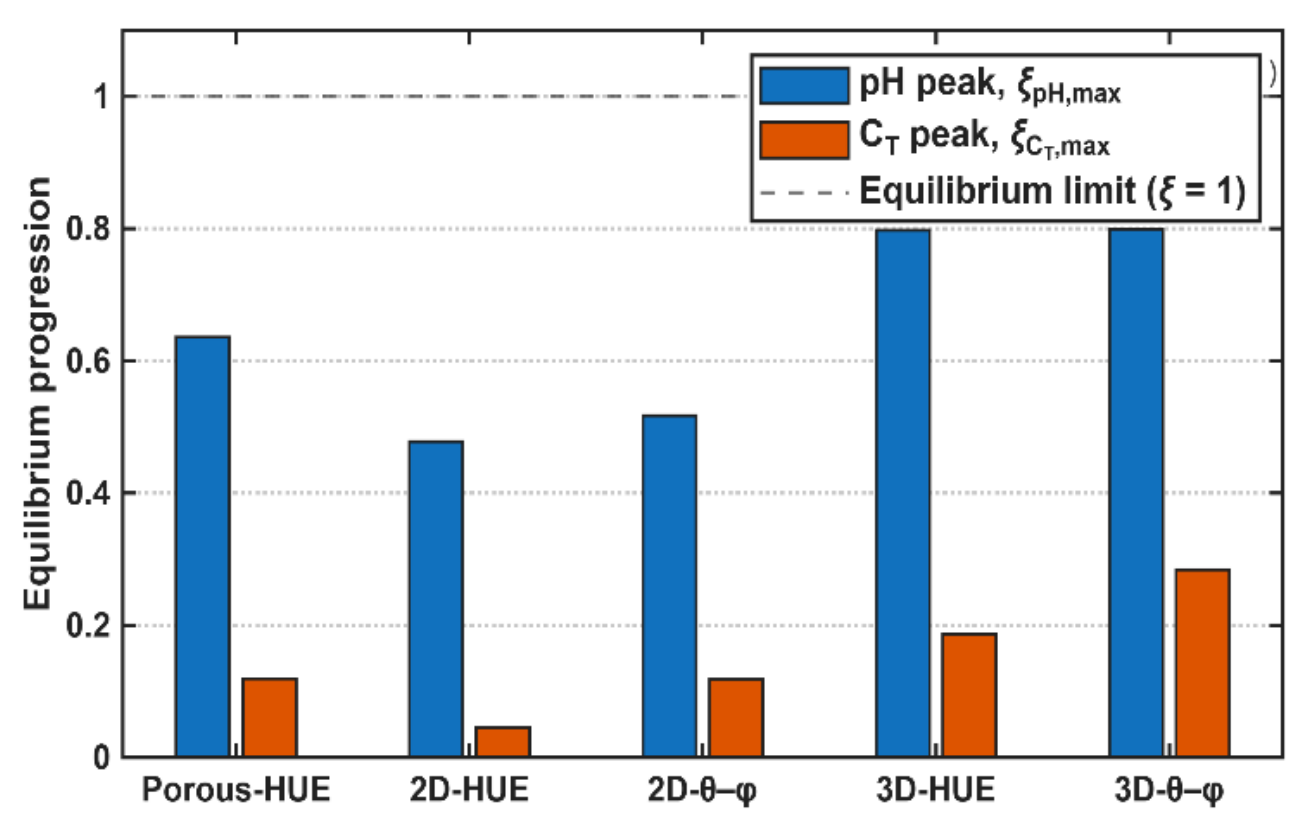

Fig 13: Maximum pH and $C_T$ equilibrium progression via convective dimensionless time (t*).

### 3.2.4 pH and $C_T$ Equilibrium Progress Scaling

The progression of the dissolution process resolved via the two colorimetric models in the different domains and geometries towards equilibrium condition ($\xi_{pH}$ $and$ $\xi_{CT}$) were scaled (Fig 13). None of the data sets fully reached the equilibrium condition.

The porous domain progressed rapidly over a short t* interval, reaching a peak of 63% of the equilibrium pH window and 12 % of $C_{sat}$ (Fig 13). The 3D-hue and 3D- θ–ϕ datasets attained 79% of equilibrium pH, and 19% and 28 % of $C_{sat,}$ respectively. The 2D-hue and θ–ϕ attained the lowest peak progression towards the equilibrium condition (47% and 51% of pHeq and 4.7% and 11% of $C_{sat}$ respectively) (Fig 13).

ξ_pH was also plotted against diffusive dimensionless time ($t^*_{diff}$) and it produced a better common trajectory than the t* (Fig S18), the normalized curves exhibit the characteristic sigmoidal behavior expected for $CO_2$ dissolution systems, consisting of an initial slow response, a period of accelerated progression, and a late-time plateau approaching equilibrium (see Fig S19) and just like the $t^*$ none of the domain reached equilibrium. The curve also showed the same peak progression trend towards the equilibrium pH and concentration seen in $t^*$ (Fig S18), and despite the comparable diffusive time scale in 2D and 3D systems, the 3D system reached much greater equilibrium-scaled progress. The results from the scaling analysis indicate that geometry and domain controlled attainable progress toward equilibrium conditions via diffusive and convective time scales however the peak $C_T$ equilibrium progression remained model dependent.

## 4 Conclusion

A constraint framework was developed to test the reliability of pH and $C_T$ measurements in spatial colorimetric mapping using scalar angular representation (hue from HSV) and bivariate angular representation (θ and ϕ) in free-fluid (2D and 3D) and quasi-2D porous domains. The results show that reliability of measurement is not governed by a single factor but by the interactions among indicator-mix composition, optical path width, color-coordinate dimensionality, and domain structure; from these, the following conclusions were drawn.

1. The indicator mix 1 with a balanced concentration ratio of BTB to TB and low MR improved both hue and θ–ϕ pH calibration curve monotonic relationships, low-pH separability, and chromatic sensitivity across 2D and 3D, whereas the mix 2 with higher dye loading of MR with BTB and TB performance depended on geometry. However, Mix 1 did not provide sufficient contrast for the porous domain pH calibration mapping under the experimental conditions due to MR's low concentration.

2. In the porous domain, the hue pH calibration curve was 33 X less susceptible to noise than θ–ϕ; however, their interpolation error (LOOCV-RMSE) was similar; therefore, interpolation error test alone should not be used to assess the reliability of pH calibration curve.

3. Baseline noise for establishing uncertainty in pH and $C_T$ measurement should be quantified directly from pH maps rather than inferred from raw RGB frames. pH maps account for all noise associated with the spatial colorimetric workflow, while RGB frames do not; a sensitive calibration curve can amplify a minor raw RGB noise to high pH noise. This situation was especially apparent in the 2D raw RGB and fixed pH map (2D θ–ϕ) spatial noise.

4. The hue colorimetric method was more reliable; its pH calibration curve was less sensitive to noise and was broadly applicable in both porous domains and free fluid. It produced consistently a low pH map noise in all domains (spatial $\sigma_{pH} = 0.066 - 0.097$), whereas θ–ϕ were more applicable in the free fluid domains.

5.The colorimetric models did not respond the same to the evolution of $CO_2$ dissolution when responsiveness was expressed via the maximum gradient of mean pH and $C_T$ vs convective dimensionless time. In 2D, the θ–ϕ model had a slightly higher gradient (pH and $C_T$) than hue, however in 3D, hue had higher pH gradient whereas the θ–ϕ had higher $C_T$ gradient.

6. The colorimetric models agree more in 2D than in 3D when represented as normalized mean pH and $C_T$ over the same convective-time progression, regardless of their absolute or maximum values.

7. Spatial dissolution maps confirmed that both models resolved the same physical plume and finger evolution, but with method-dependent sharpness: θ–ϕ produced sharper plume boundaries and stronger gradient contrast, while hue produced smoother, more spatially stable background fields. Geometry mainly controlled the convective transport (3D yielding earlier, deeper, broader fingers than 2D), and inclusion of porous media increased the persistence of pore-scale heterogeneity.

Overall, these results demonstrate that colorimetric methods can be deployed to measure convective dissolution processes in a range of domains and geometries, but care is required to determine the most appropriate pH quantification route.

**Acknowledgements**

The authors acknowledge financial support from the University of Tennessee (UTK) Institute for a Secure and Sustainable Environment (ISSE).

# Supplementary Material for:
# Spatial Colorimetric Mapping of pH and Dissolved Carbon during Convective Dissolution of $CO_2$: Constraints from Domain, Geometry, and Indicator Mix on Measurement Reliability

Joe-sam Nkuah[1], Vance Sisson[2], Haochen Li[1], Anna Herring[1]
1. Department of Civil and Environmental Engineering, University of Tennessee–Knoxville; TN, USA
2. Department of Chemical Engineering, University of Tennessee–Knoxville; TN, USA

## 1.0 Image Processing

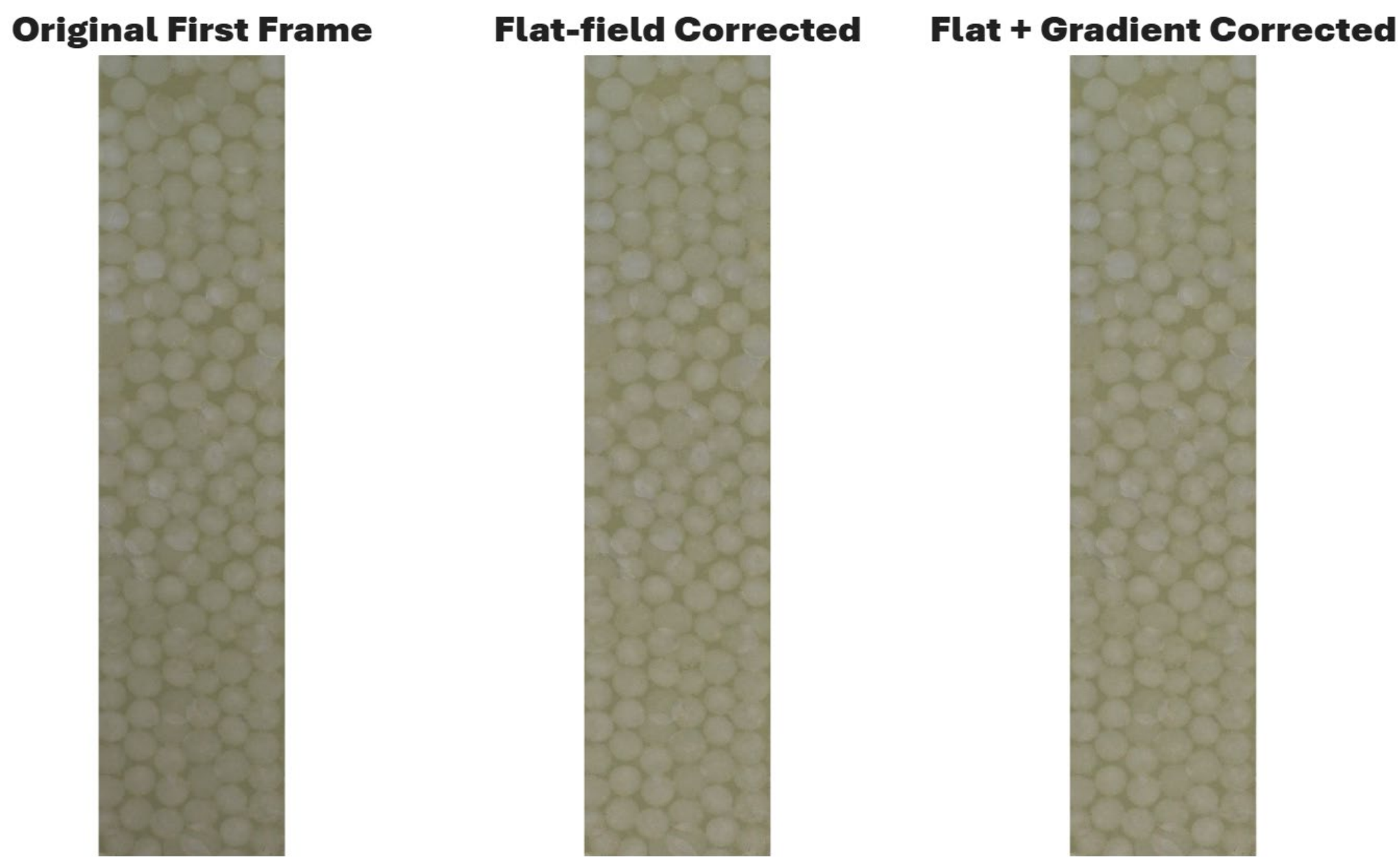

Fig S1: Image preprocessing workflow for the porous-domain reference frame.

## 2.0 Pore Segmentation

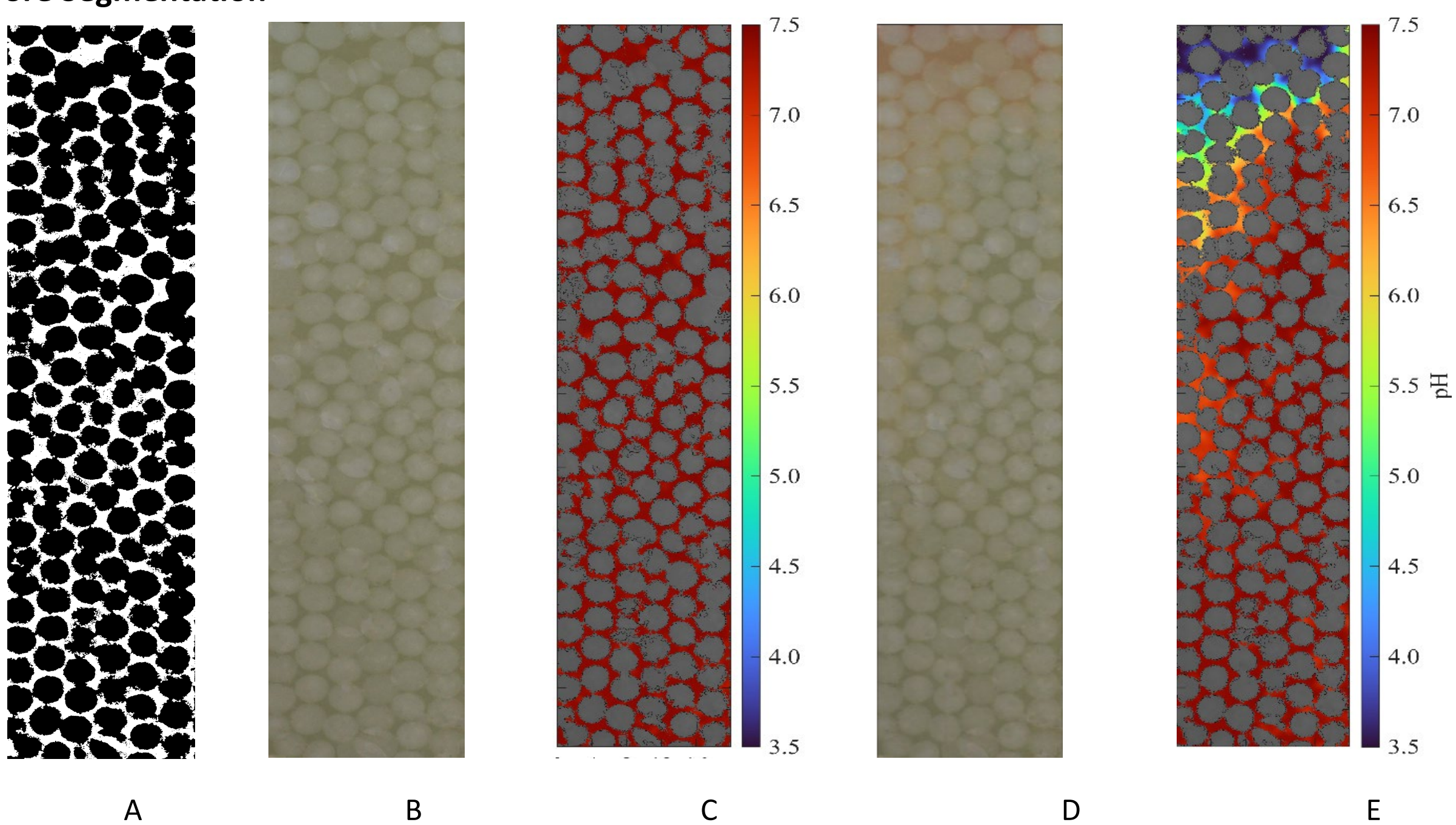

Fig S2: . Porous-domain segmentation pore mask used for pH and $C_T$ mapping.
A: segmented pore mask, B& D: raw frames and C&E : mapped frames

## 3.0 Impact of indicator mix concentration on pH calibration

Two different indicator mix concentrations (mix-1 (13.2 BTB / 13.2 TB / 6.61 MR) mg/l and mix-2 (50 BTB / 25 TB / 32 MR) mg/l) were used to prepare two sets of calibration solutions with pH ranging from 3.5 to 8. The solution was imaged and processed in 2D and 3D cells, and the resulting calibration curves were evaluated to determine the effect of dye loading on optical width and its impact on the hue and θ–ϕ models.

The curves were evaluated by measuring the separability of points in the lower pH range (4-5.5) using the pH gradients dH/dpH, dθ/dpH, and dϕ/dpH. This analysis was done to evaluate how geometry and dye loading impact the ability of the colorimetric models to resolve low pH, which is very relevant to $CO_2$ dissolution. Regression ($R^2$) was also used to evaluate the overall relationship to pH. The results show that indicator mix concentration and optical width strongly influenced the sensitivity of the chromatic pH calibration. Overall, the balanced – low mix-1 solution (13.2 BTB / 13.2 TB / 6.61 MR) produced the most stable calibration behavior across all free fluid domain, maintaining smooth monotonic hue evolution with good lower-pH range separability in the 2D and 3D cells, whereas the mix-2 (50 BTB / 25 TB / 32 MR) behavior was strongly dependent on geometry (Fig S7. and table S1). Mix-1 maintained stronger lower-pH θ–ϕ color separability in both 2D and 3D, while the reduced optical width of the 2D cell partially restored θ and ϕ separability even for the higher-concentration mix-2 (Table S2- S3). In general, mix-2 performed well in 2D across the range of pH investigated for both hue and θ–ϕ quantification routes. However, calibration curves for the 3D geometry exhibited compression in the low-pH range (as quantified by calibration gradients dH/dpH, dθ/dpH, and dϕ/dpH approaching 0 at pH < 5.5 (figs. S3-S7 and table S1, S2 and S3).

Table S1: Impact of indicator mix concentration and geometry on hue pH calibration curve

| Mix | Geometry | Hue $R^2$ | Lower pH Separability (5.5 -4) dH/dpH |
|---|---|---|---|
| Mix-1 (13.2 BTB / 13.2 TB / 6.61 MR) | 3D | 0.989 | 0.022 |
| Mix-2 (50 BTB / 25 TB / 32 MR) | 3D | 0.920 | 0.008 |
| Mix-1 | 2D | 0.997 | 0.028 |
| Mix-2 | 2D | 0.987 | 0.025 |

Table S2: Impact of indicator mix concentration and geometry on θ–coordinate pH calibration response

| Mix | Geometry | θ $R^2$ | Lower pH Separability, (5.5 - 4) dθ/dpH |
|---|---|---|---|
| Mix-1 | 3D | 0.992 | 0.020 |
| Mix-2 | 3D | 0.979 | 0.005 |
| Mix-1 | 2D | 0.995 | 0.090 |
| Mix-2 | 2D | 0.974 | 0.018 |

Table S3: Impact of indicator mix concentration and geometry on ϕ coordinate pH calibration response

| Mix | Geometry | ϕ $R^2$ | Low pH Separability (5.5 - 4), dϕ/dpH |
|---|---|---|---|
| Mix-1 | 3D | 0.986 | 0.086 |
| Mix-2 | 3D | 0.928 | 0.001 |
| Mix-1 | 2D | 0.984 | 0.018 |
| Mix-2 | 2D | 0.968 | 0.019 |

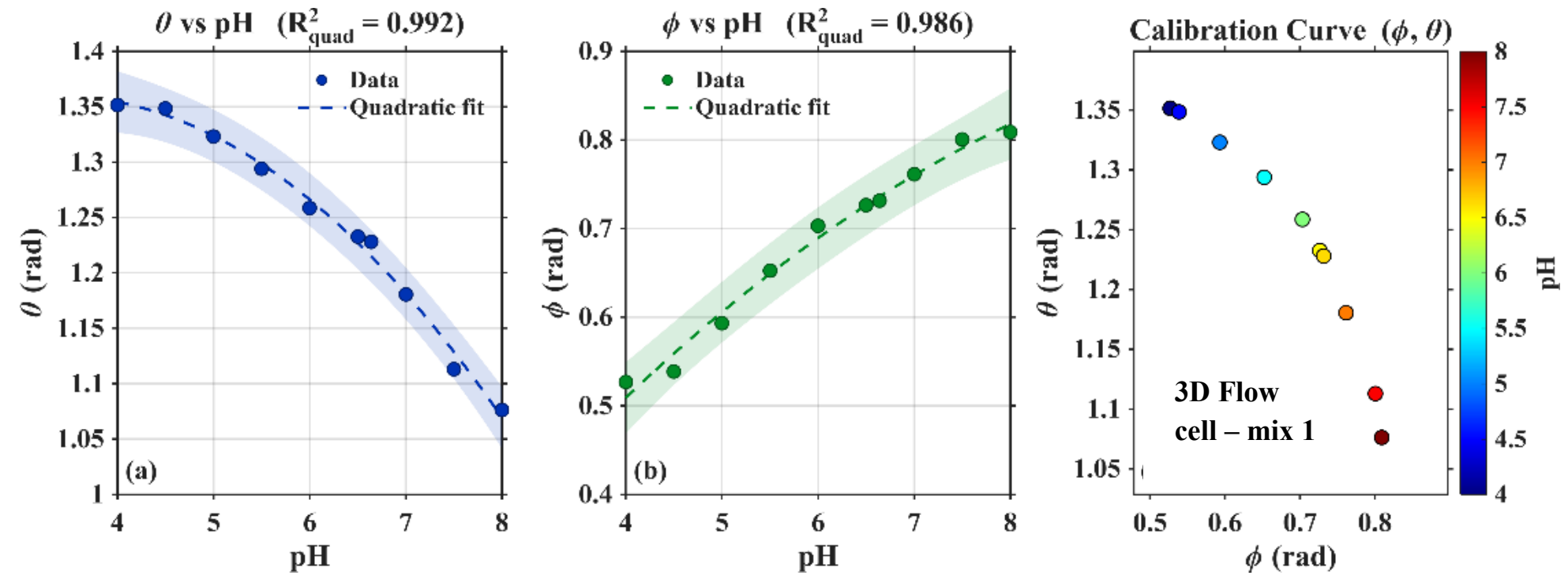

Fig S3: θ–ϕ pH calibration curve for the 3D free-fluid cell using Mix 1.

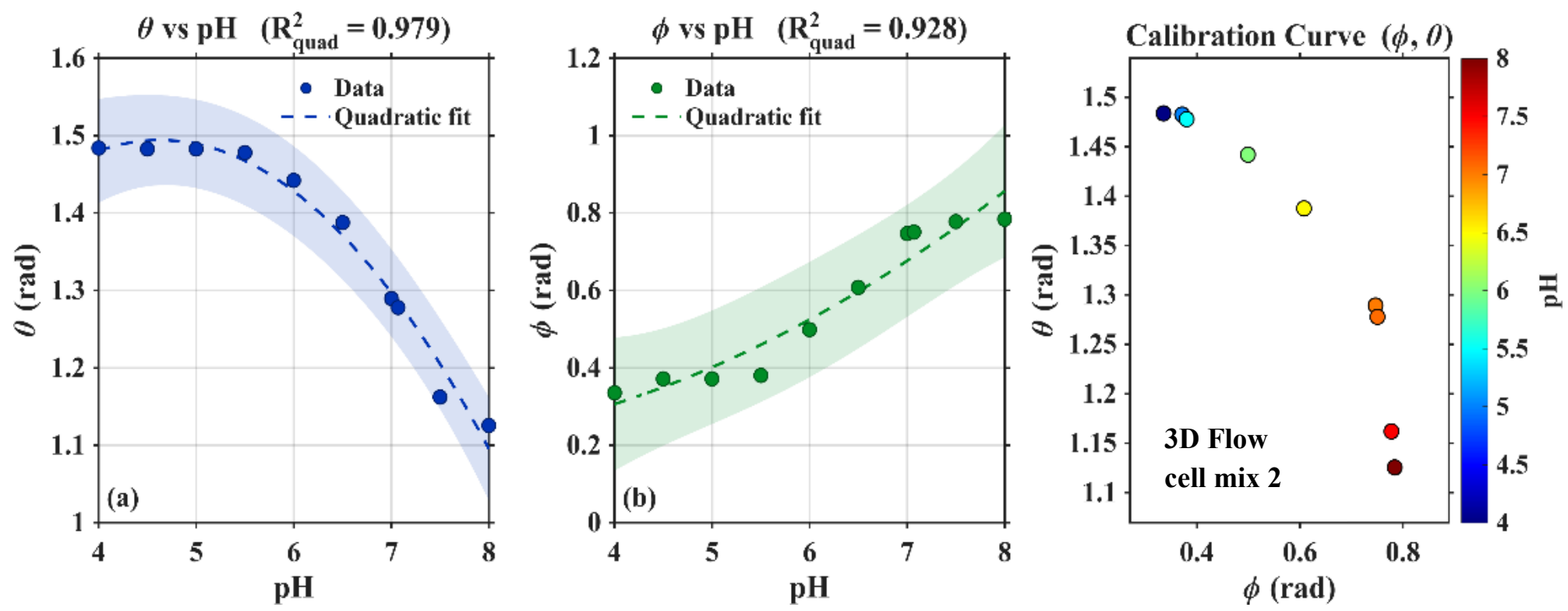

Fig S4: θ–ϕ pH calibration curve for the 3D free-fluid cell using mix 2.

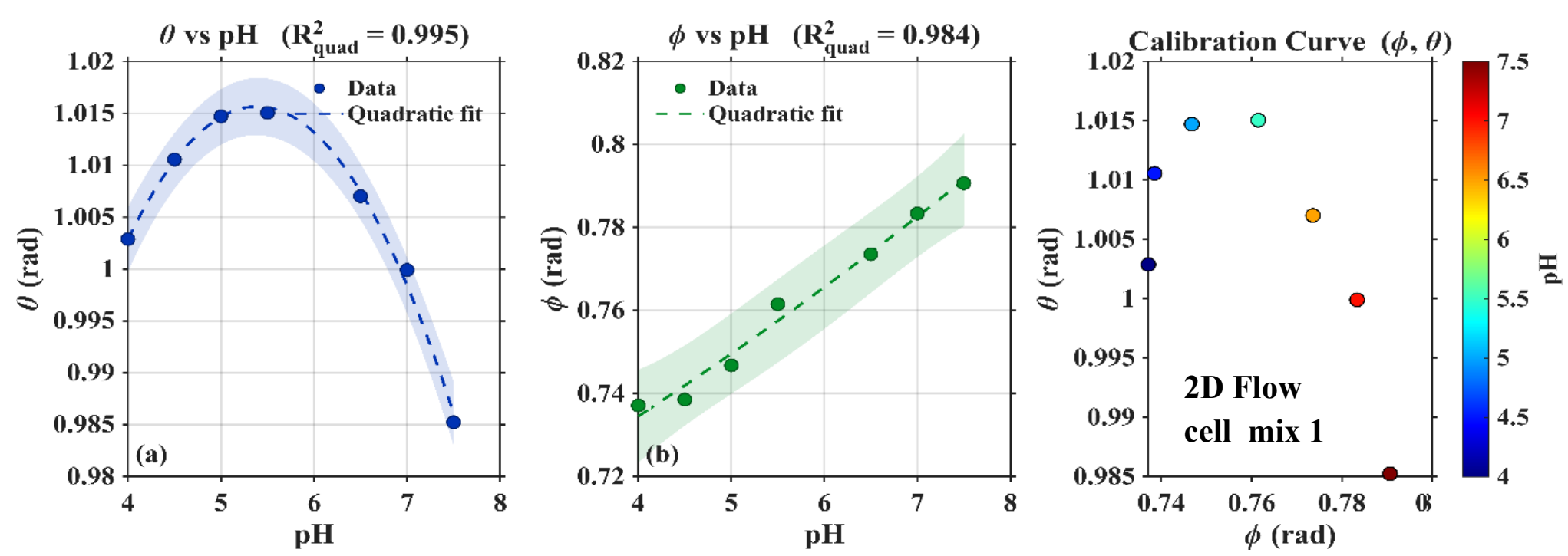

Fig S5: θ–ϕ pH calibration curve for the 2D free-fluid cell using Mix 1.

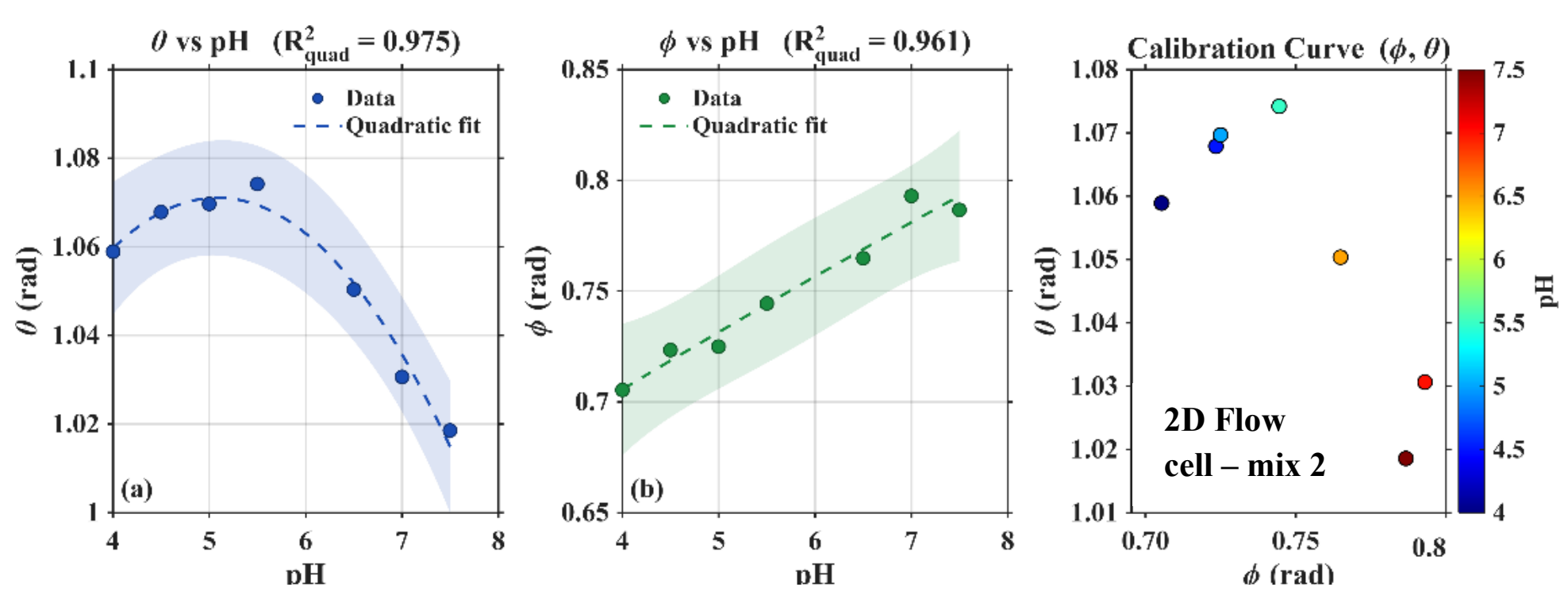

Fig S6: θ–ϕ pH calibration curve for the 2D free-fluid cell using mix 2.

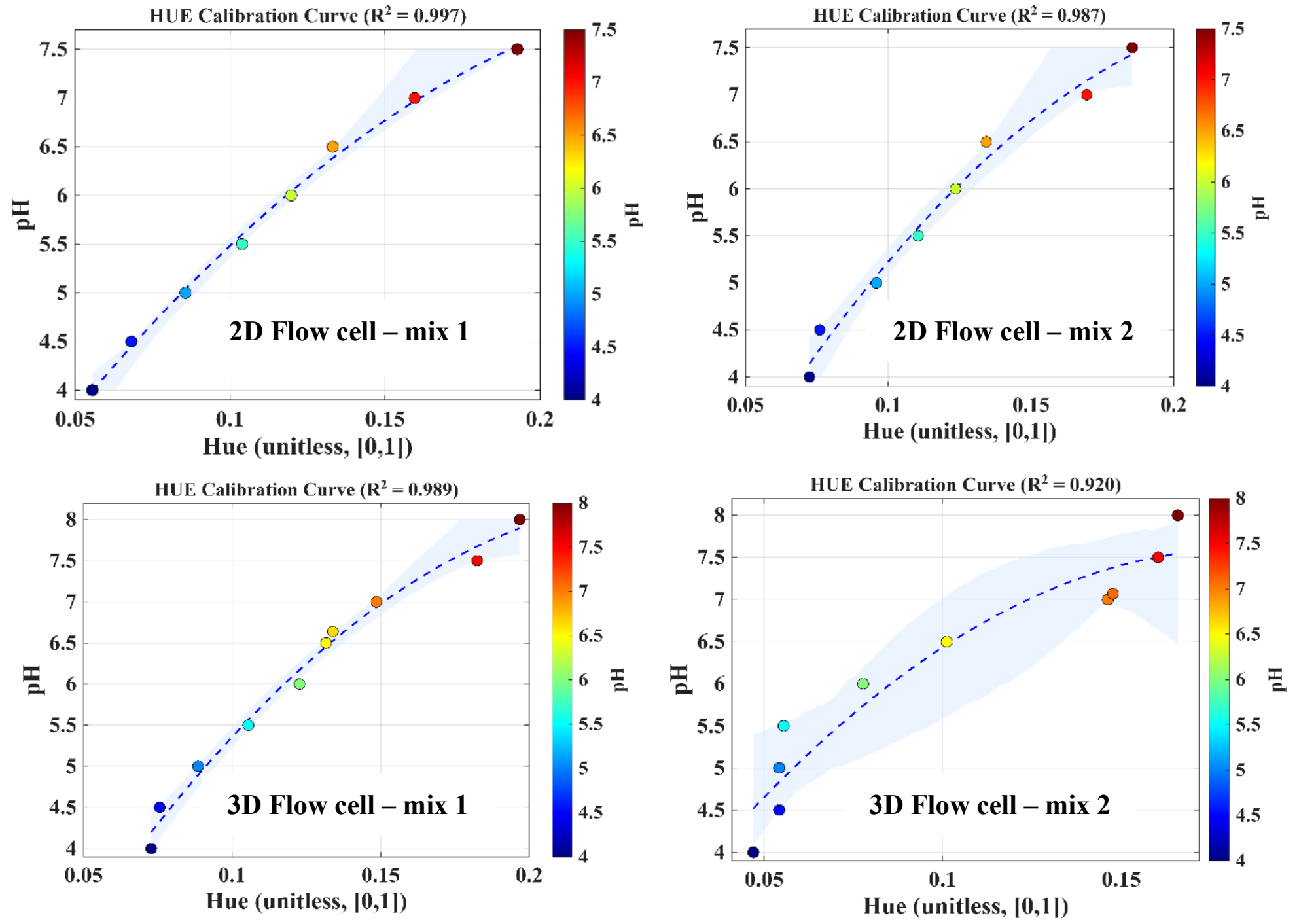

Fig S7: Hue-based pH calibration curves for mix 1 and mix 2 across free-fluid geometries. Calibration curves show the relationship between the normalized hue, H ∈ [0, 1], and the measured pH for the 2D free-fluid.

## 4.0 Stability Test (temporal and spatial drift test)

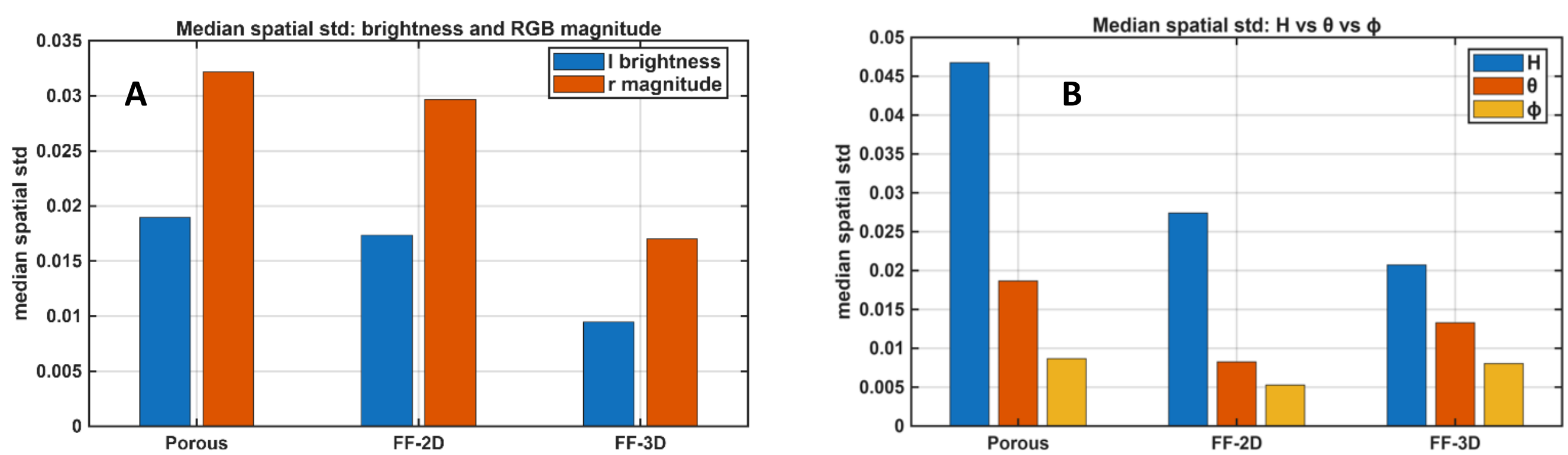

Fig S8: Raw fixed-pH frame stability (A) spatial variability in brightness and RGB-vector magnitude,(B) spatial variability in raw hue, θ and ϕ

Table S4: Fixed-pH map stability metrics for hue and θ–ϕ models across domains.

| Metric | Porous HUE (quasi 2D) | 2D HUE | 2D θ–ϕ | 3D HUE | 3D θ–ϕ |
|---|---|---|---|---|---|
| pH_true | 7.22 | 7.22 | 7.22 | 7.22 | 7.22 |
| Mean pH | 7.217 | 7.028 | 6.938 | 7.195 | 7.215 |
| Bias (pH) | 0.003 | 0.192 | 0.282 | 0.025 | 0.005 |
| Temporal σ_pH | 0.0015 | 0.0062 | 0.010 | 0.0031 | 0.0041 |
| Median spatial σ_pH | 0.0970 | 0.0880 | 0.1500 | 0.0660 | 0.0370 |
| P95-P5 Range | 0.2973 | 0.2823 | 0.5051 | 0.2030 | 0.1285 |
| P95 Pixel Temporal σ | 0.085 | 0.123 | 0.268 | 0.044 | 0.064 |
| Drift (pH/min) | $-7 \times 10^{-7}$ | $-8.8 \times 10^{-5}$ | $-2 \times 10^{-4}$ | $-8.5 \times 10^{-6}$ | $1.2 \times 10^{-7}$ |

## 5.0 pH noise propagation

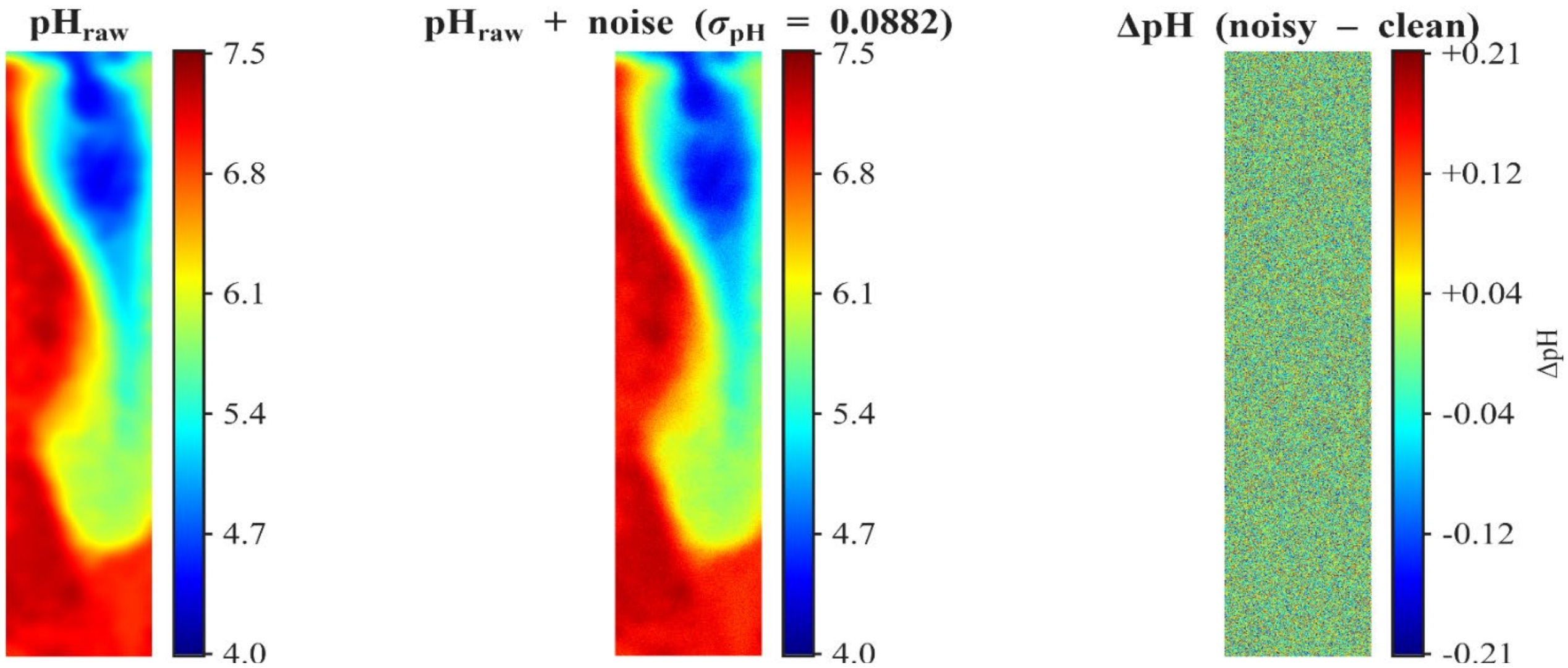

Fig S9: Monte Carlo pH-noise propagation for the 2D free-fluid hue map.

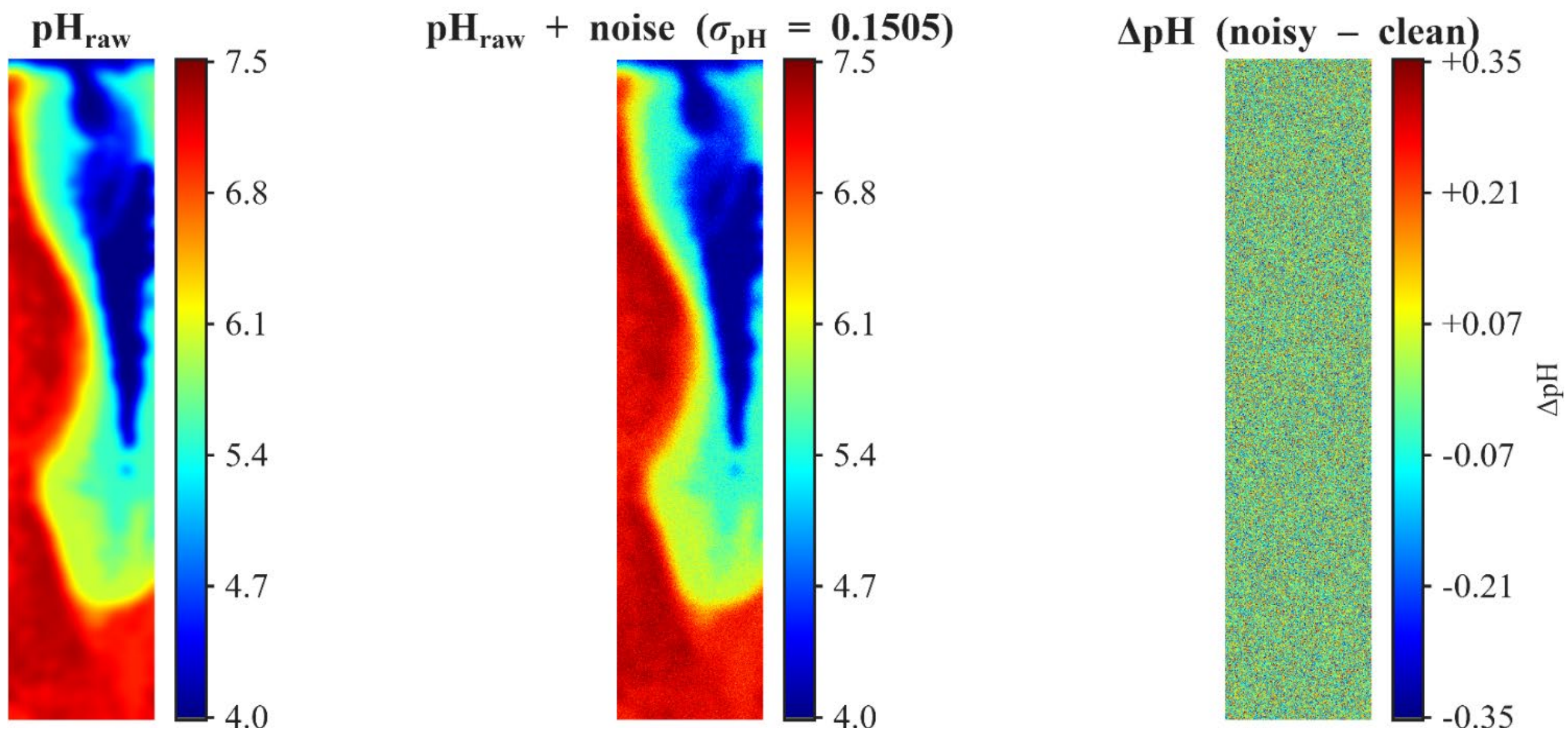

Fig S10: Monte Carlo pH-noise propagation for the 2D free-fluid θ–ϕ map

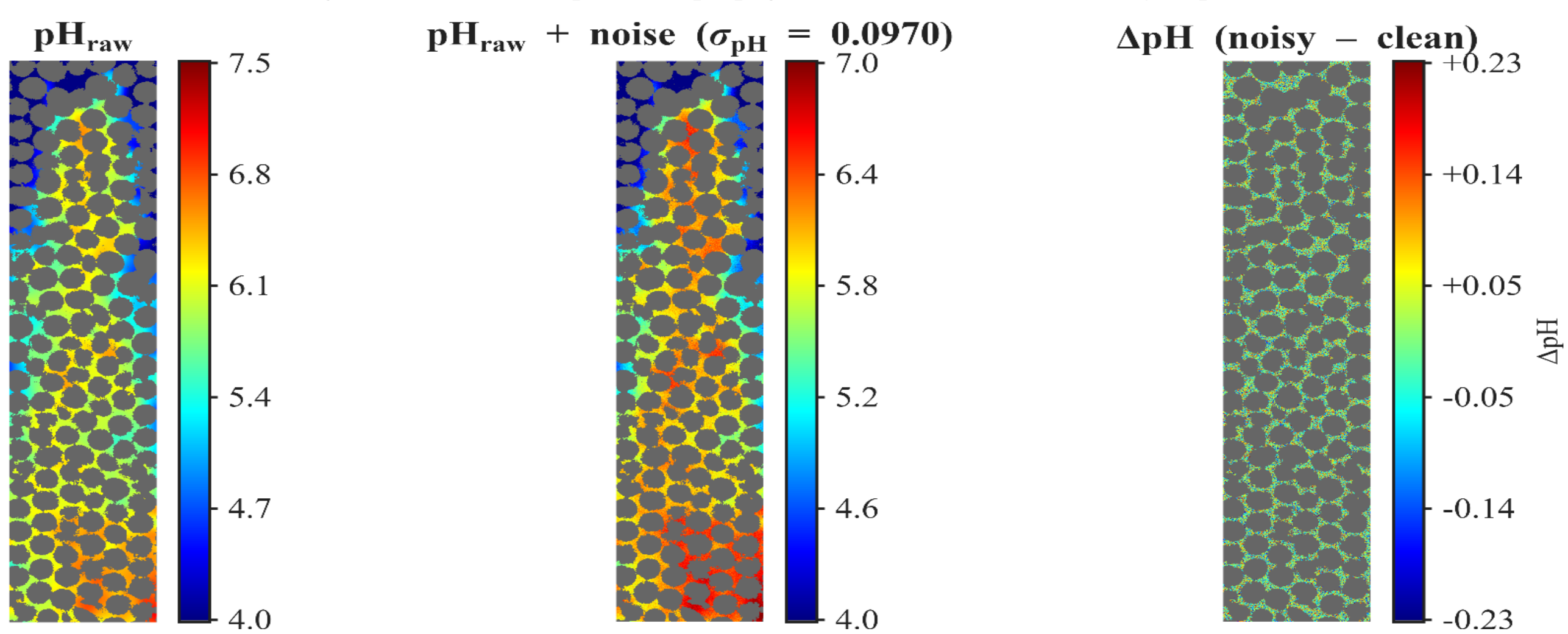

Fig S11: Monte Carlo pH-noise propagation for the porous-domain hue map

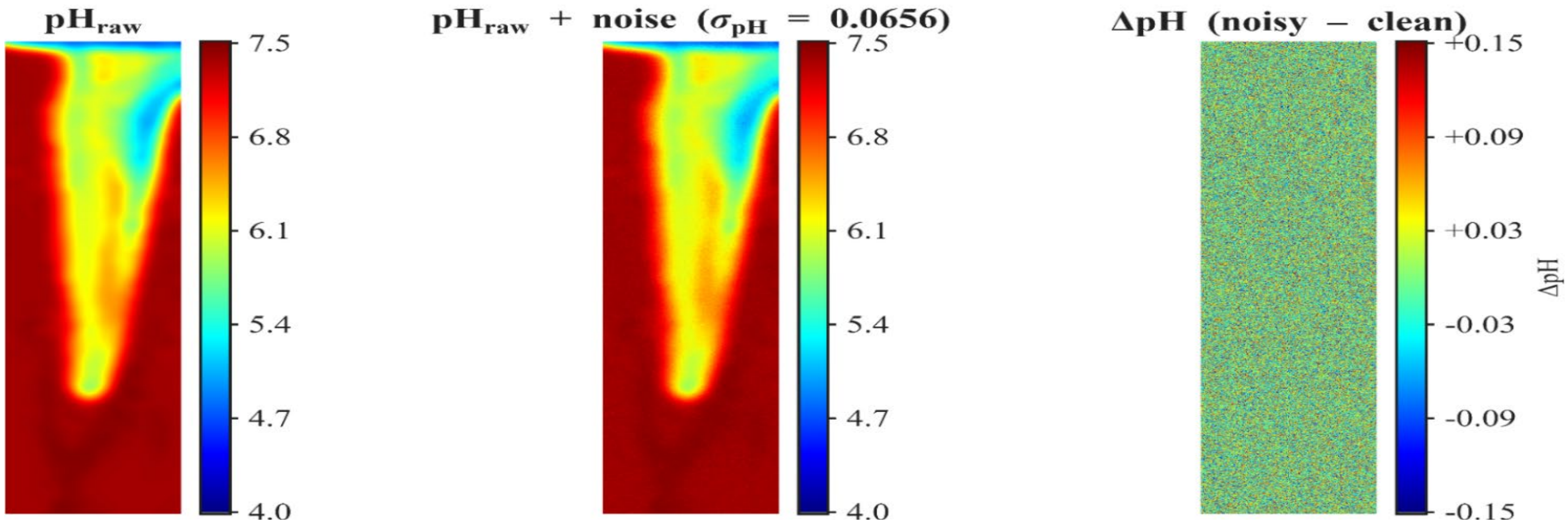

Fig S12: Monte Carlo pH-noise propagation for the 3D free-fluid hue map.

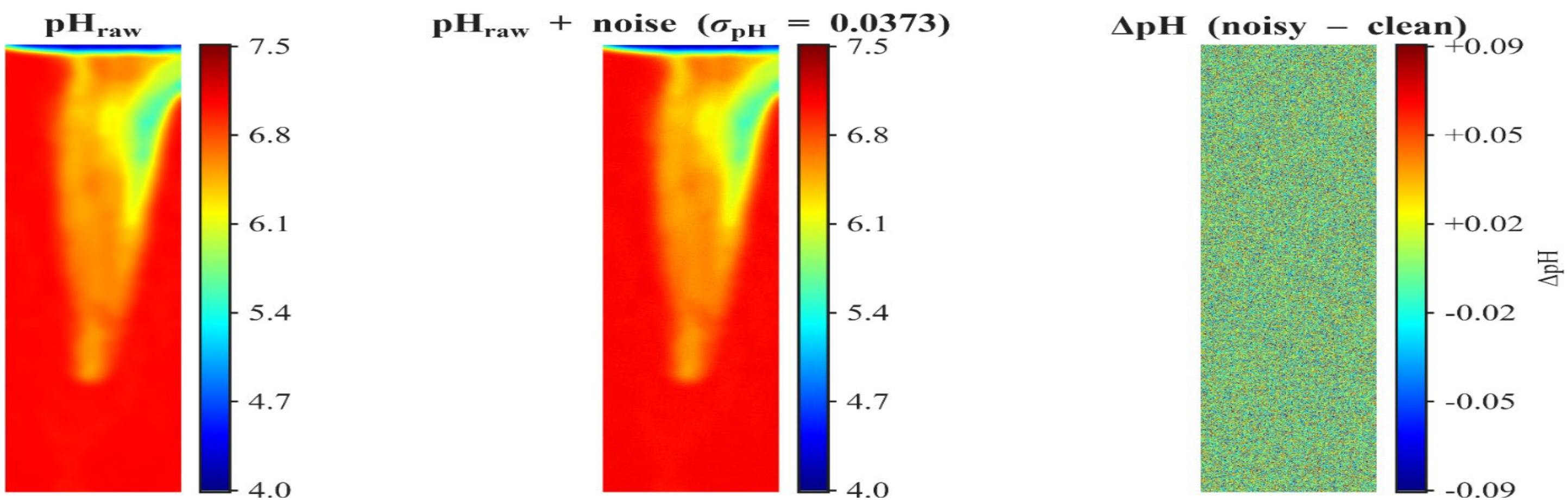

Fig S13: Monte Carlo pH-noise propagation for the 3D free-fluid θ–ϕ map.

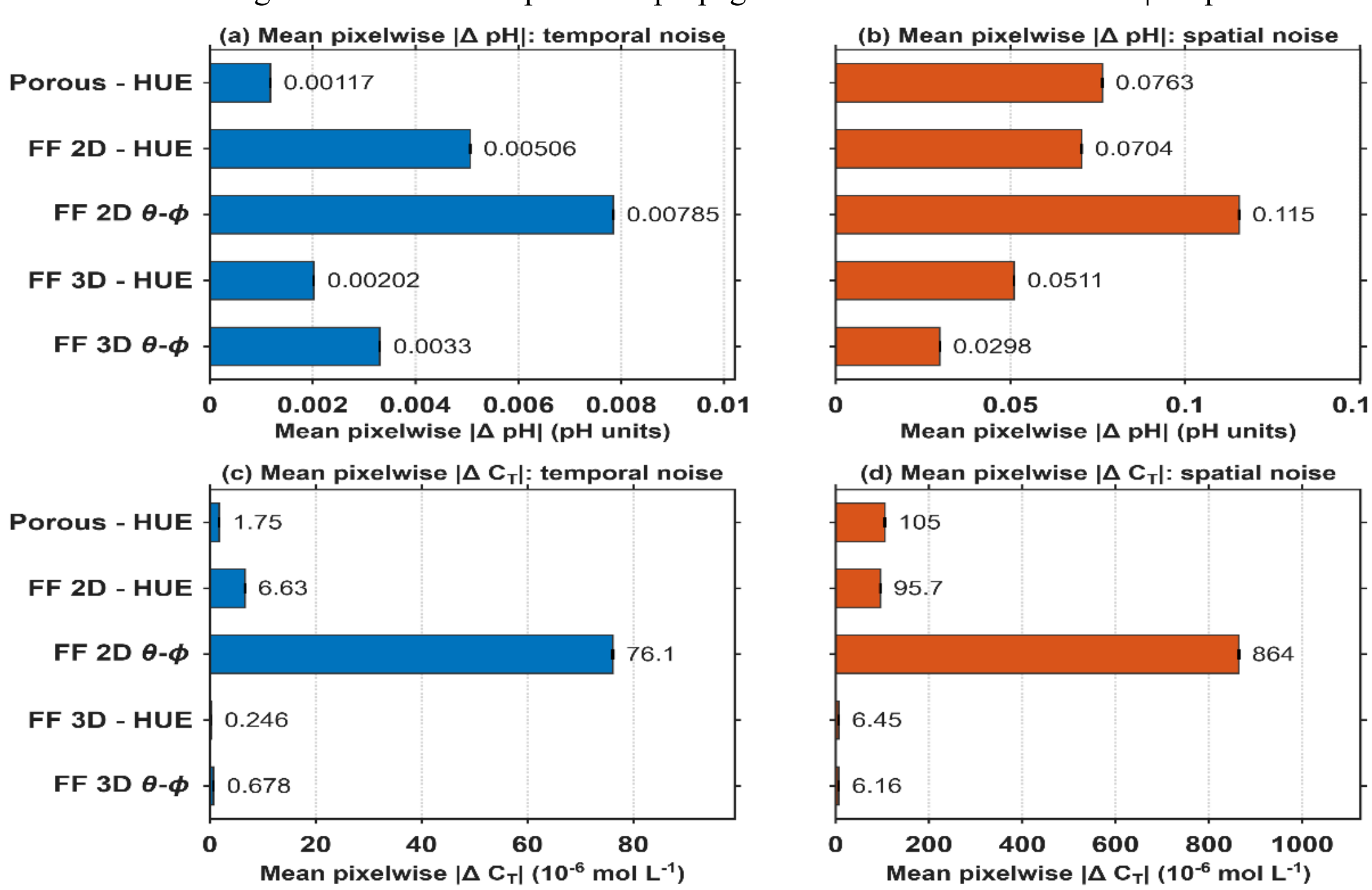

Fig S 14: Pixelwise changes in pH and $C_t$ after noise propagation. Mean absolute pixelwise changes in pH under (a) temporal and (b) spatial noise, and in $C_t$ under (c) temporal and (d) spatial noise. bars show the mean of 100 Monte Carlo realizations; error bars show the standard deviation across realizations.

## 6.0 Dimensionless Time Scaling of Mean pH and $C_T$ Curves

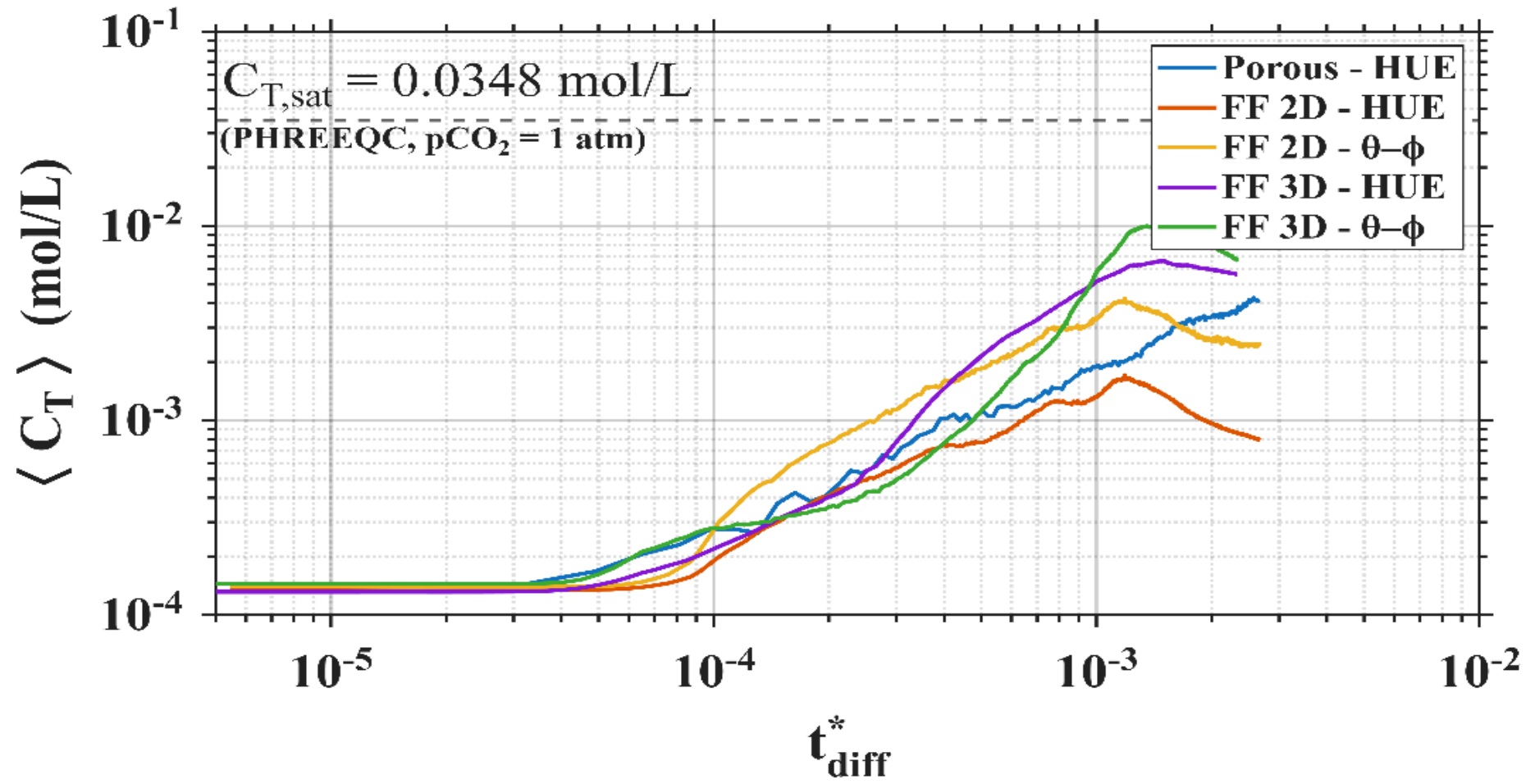

Fig S15: Diffusive dimensionless time scaling of mean $C_T$

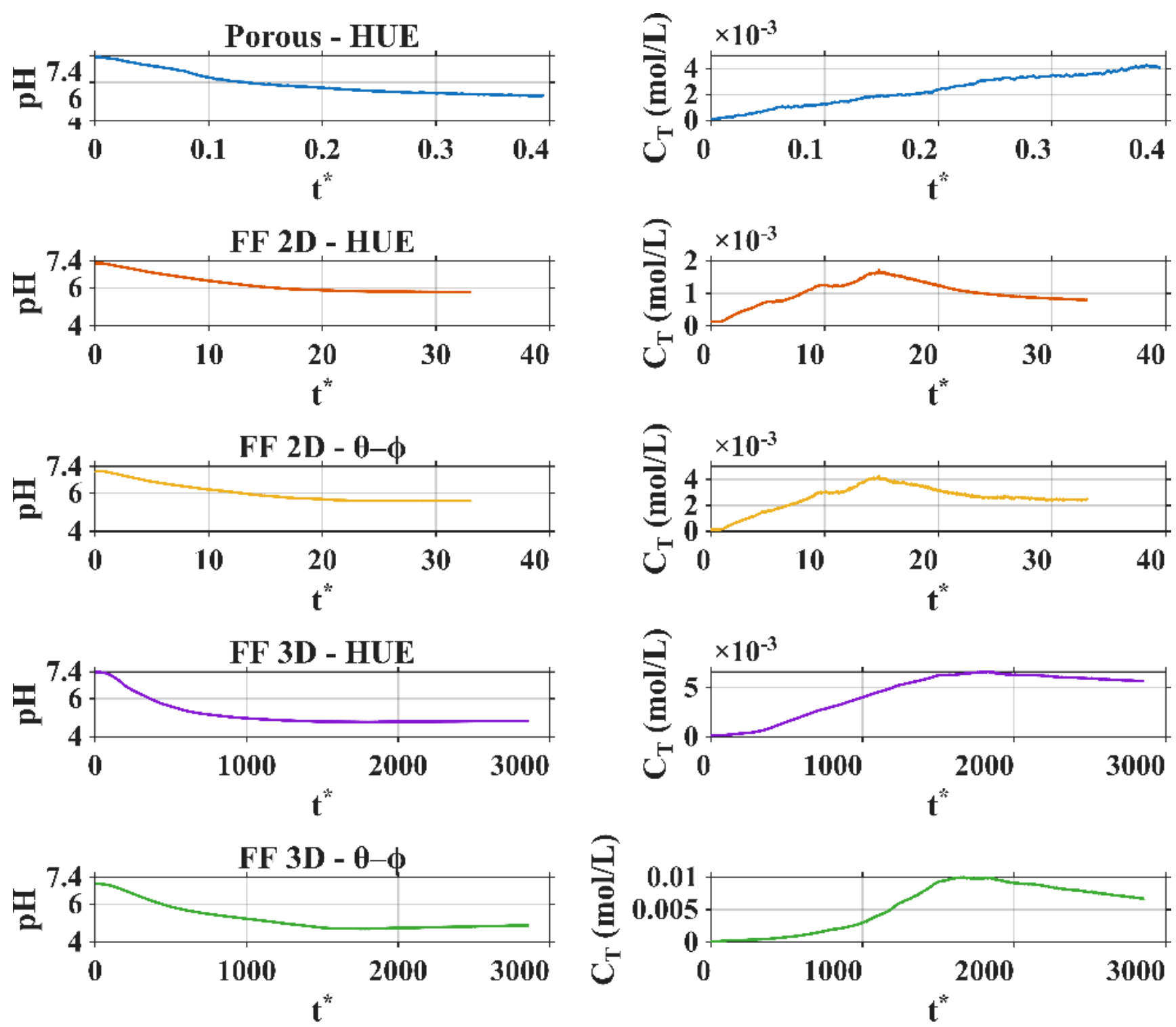

Fig S16: Convective dimensionless-time scaling of mean pH and $C_T$ curves

## 7.0 pH and $C_T$ Evolution Curves Shape Agreement and Gradient Metrics

Table S5: Peak temporal-gradient response and late-time stabilization of Mean pH and $C_T$ curves

| Dataset | Max Temporal Response | | Late-Time Stabilization | |
|---|---|---|---|---|
| | \|dpH/dt*\| | \|dC_T/dt*\| | dpH/dt* | dC_T/dt* |
| Porous-HUE | 19.709606 | 0.067610 | -4.704 | 0.009557 |
| 2D-HUE | 0.236902 | 0.000136 | -0.00228 | 2.369e-06 |
| 2D- θ-ϕ | 0.322267 | 0.000339 | -0.00258 | 5.609e-06 |
| 3D-HUE | 0.007346 | 5.927e-06 | 4.785e-05 | -5.72E-07 |
| 3D- θ-ϕ | 0.003654 | 1.942e-05 | 0.00016 | -1.80E-06 |

Note: The porous hue temporal response results in table S5 is high because of short range of convective dimensionless time (t*) as compared to free fluid dataset, so it should not be interpreted directly as having greater physical response than the free fluid dataset.

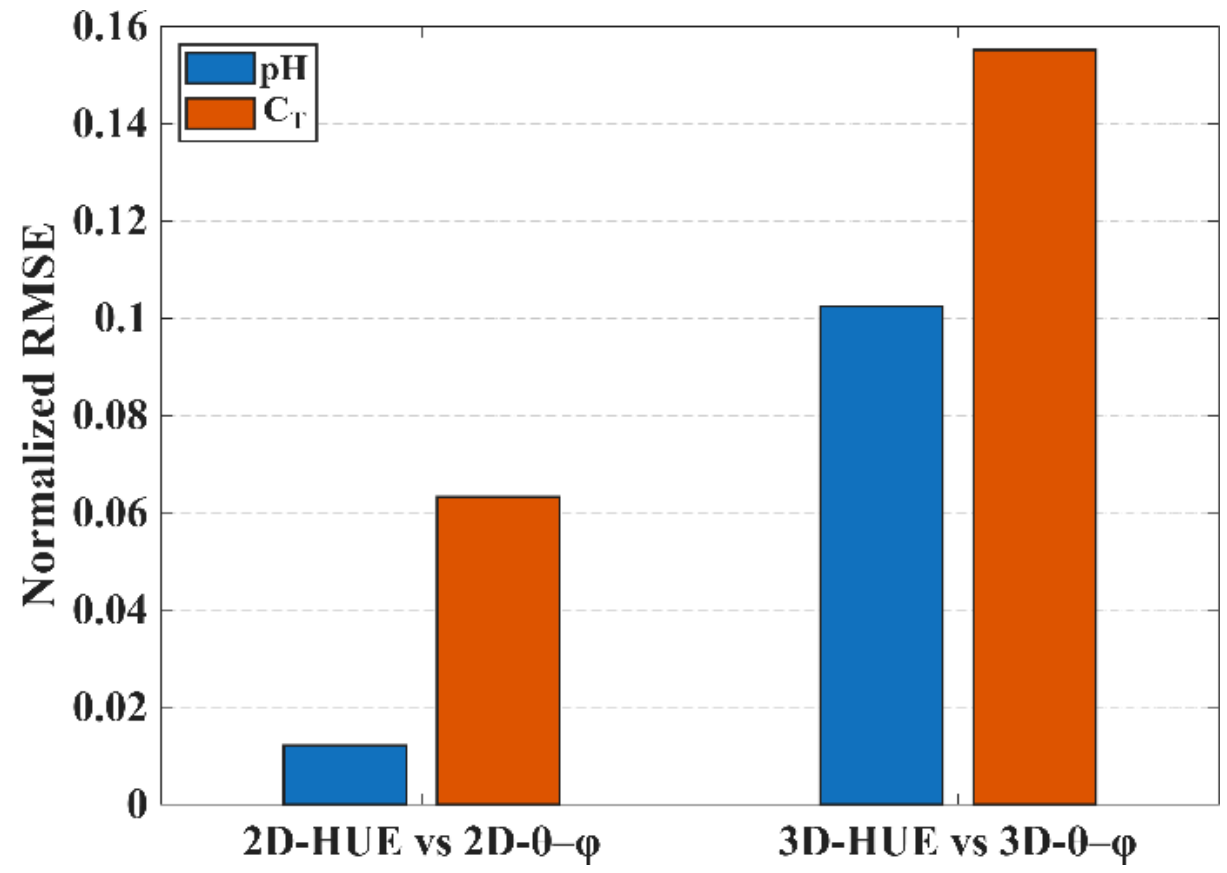

Fig S17: Normalized RMSE of hue and θ–φ curve-shape agreement

## 8.0 pH and $C_T$ Equilibrium Progress Scaling

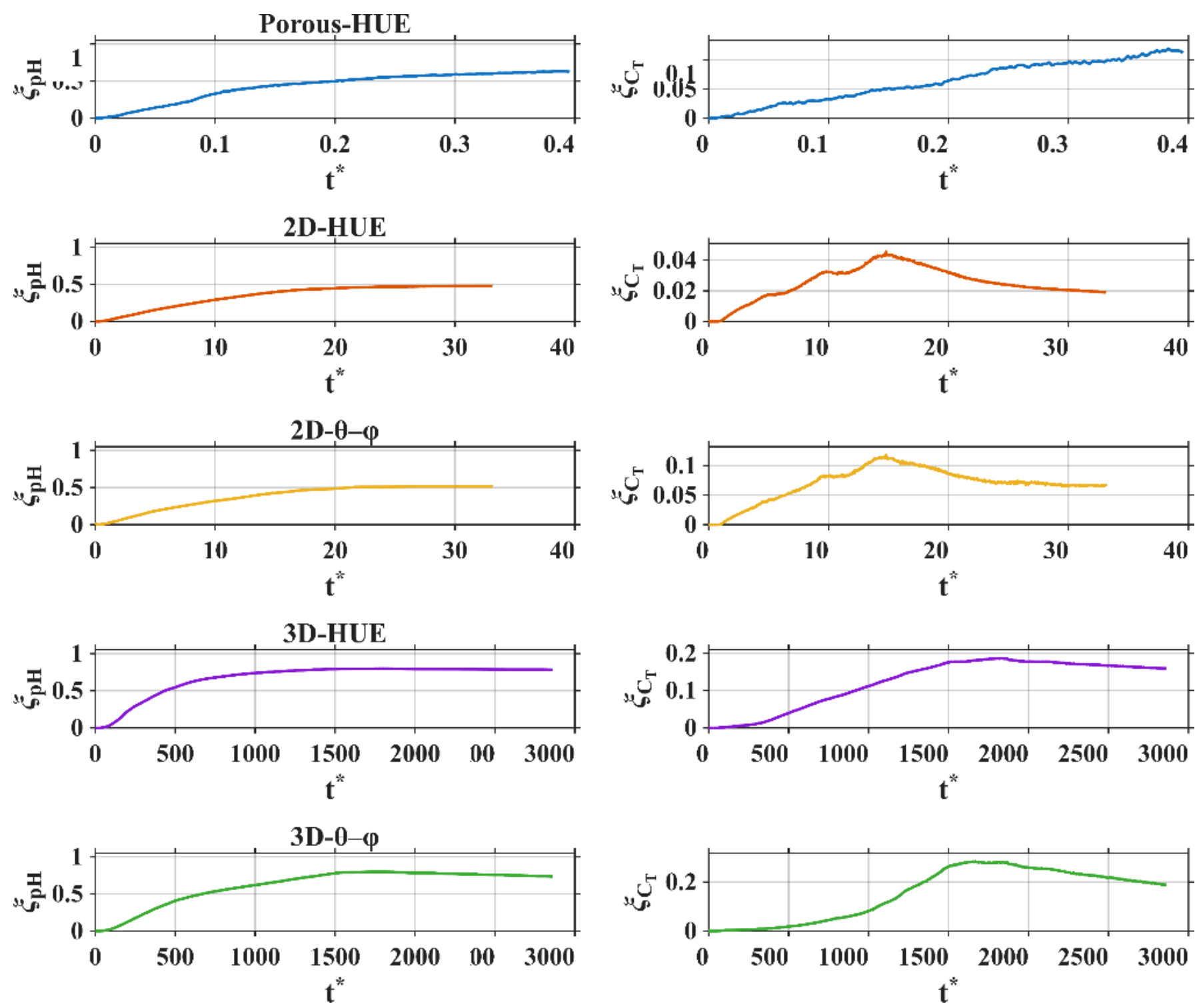

Fig S18: Equilibrium progress scaling of pH and $C_T$ curves via convective dimensionless time

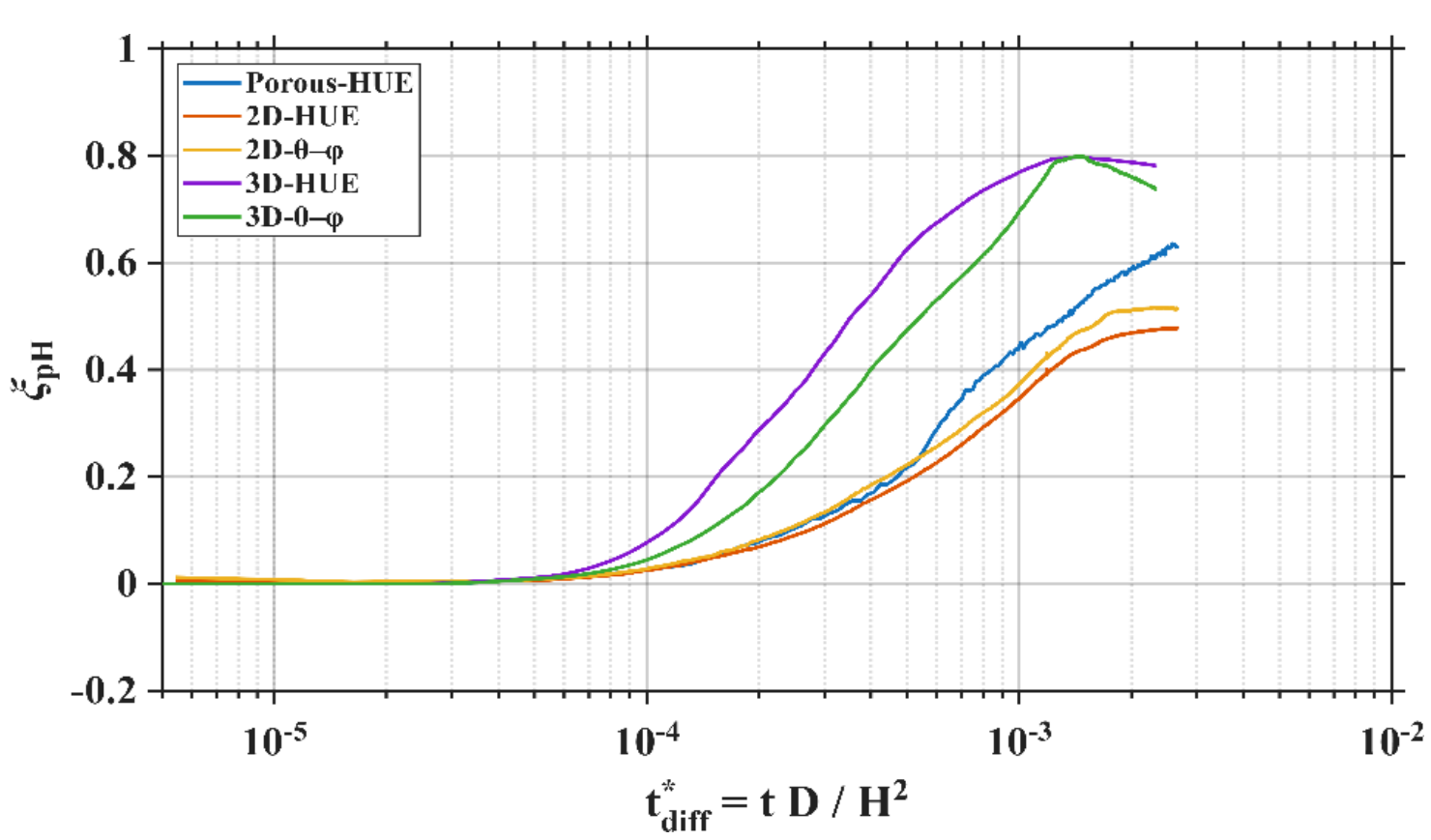

Fig S19: Equilibrium progress scaling of pH curves via diffusive dimensionless time.

### 9.0 Mean pH and $C_T$ ( Stable diffusive layer )

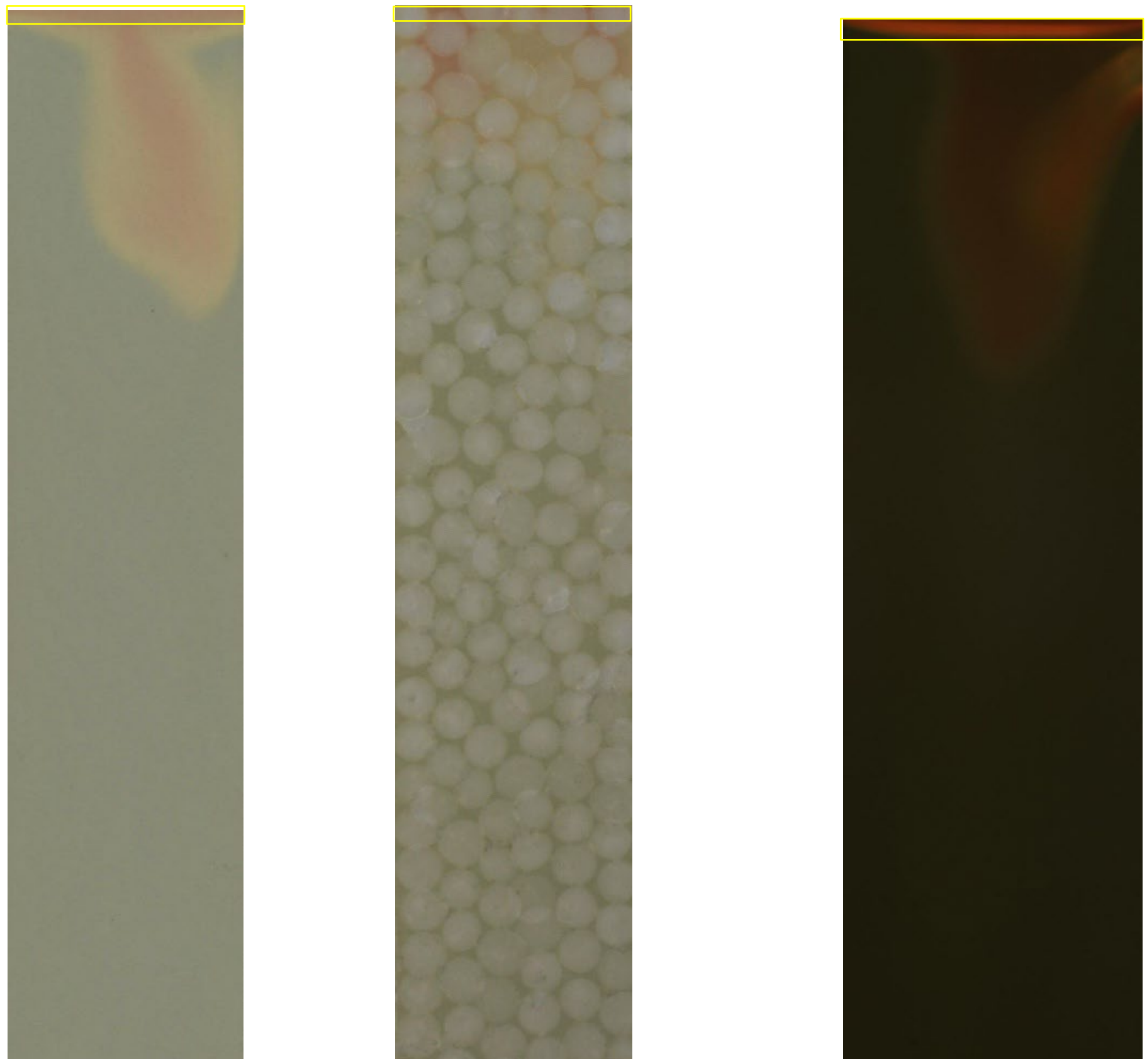

A: 2D B: Porous domain C: 3D

Fig S20: $CO_2$ dissolution frames with a stable diffusive layer in the yellow region at the top. All mean pH and $C_T$ analyses were performed below the yellow region (stable diffusive layer).

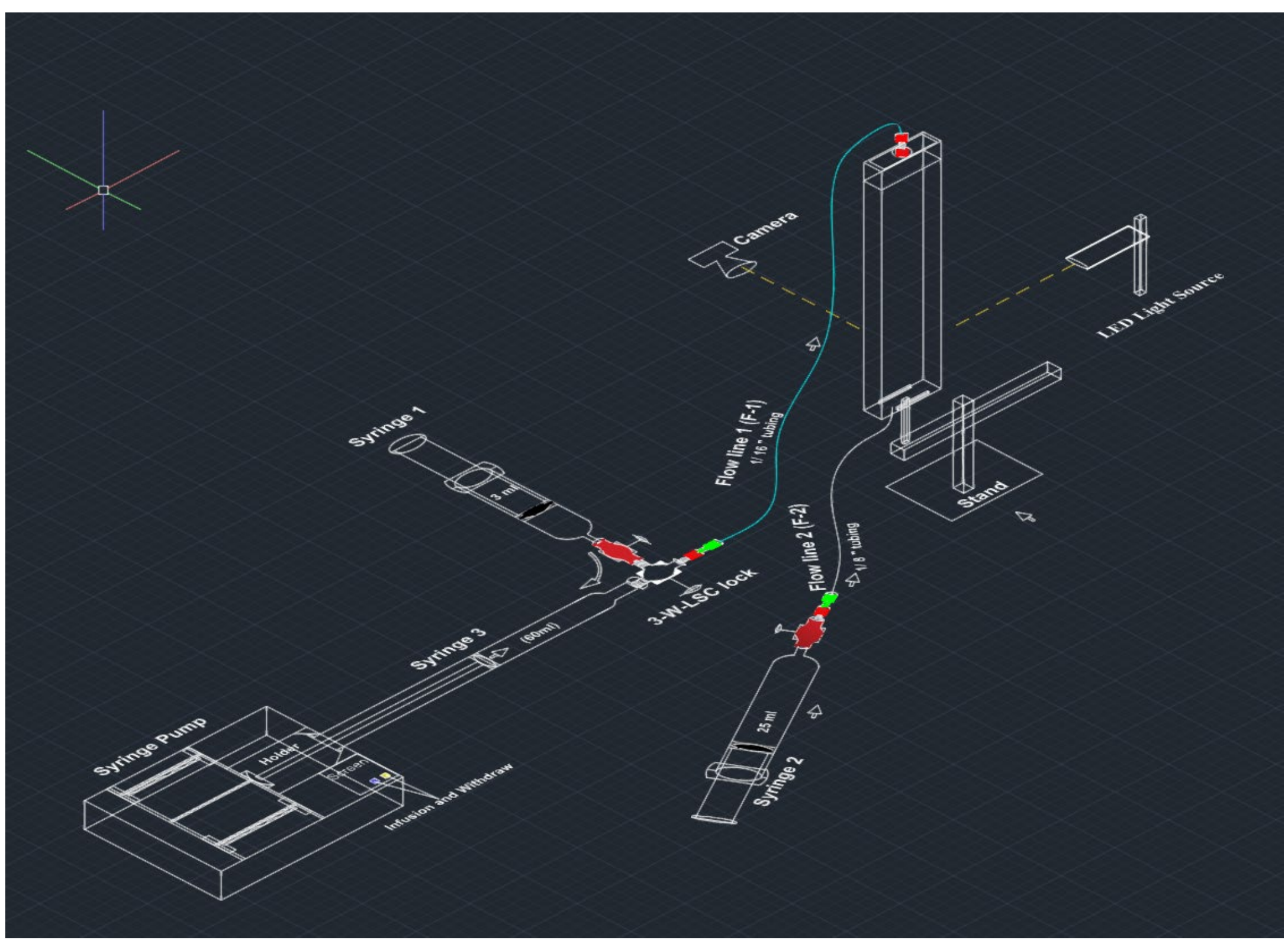

Fig S21: AutoCAD drawing of Experimental setup